\documentclass[a4paper,titlepage]{article}

\usepackage{moreverb}

\usepackage[colorlinks,bookmarksopen,bookmarksnumbered,citecolor=red,urlcolor=red]{hyperref}

\usepackage[svgnames, x11names]{xcolor}
\definecolor{RED}{rgb}{0.8, 0.0, 0.0}
\definecolor{MAGENTA}{rgb}{0.55, 0.0, 0.55}
\definecolor{ORANGE}{rgb}{0.8, 0.33, 0.0}
\definecolor{BLUE}{rgb}{0.0, 0.0, 0.8}

\usepackage[normalem]{ulem} 

\newcommand{\modc}[1]{#1}

\usepackage{listings}

\lstdefinelanguage{XML}
{
basicstyle=\ttfamily\footnotesize,
  morestring=[b]",
  moredelim=[s][\bfseries\color{Maroon}]{<}{\ },
  moredelim=[s][\bfseries\color{Maroon}]{</}{>},
  moredelim=[l][\bfseries\color{Maroon}]{/>},
  moredelim=[l][\bfseries\color{Maroon}]{>},
  morecomment=[s]{<?}{?>},
  morecomment=[s]{<!--}{-->},
  commentstyle=\color{gray},
  stringstyle=\color{blue},
  identifierstyle=\color{red}
}
\usepackage[pdftex]{graphicx}
\graphicspath{{./../}}
\DeclareGraphicsExtensions{.pdf}

\usepackage{amssymb}
\usepackage{mathtools}

\usepackage[caption=false,font=scriptsize]{subfig}

\usepackage{algorithmicx}
\usepackage{algpseudocode}
\usepackage[ruled]{algorithm}
\definecolor{light-gray}{gray}{0.75}
\algrenewcommand{\algorithmiccomment}[1]{\hskip3em{{\footnotesize \textcolor{light-gray}{$\blacktriangleright$}}} #1}

\usepackage{multirow}

\usepackage{hyperref} 

\usepackage{xspace}

\usepackage{enumitem}

\begin{document}

\title{Towards a Data-driven IoT Software Architecture for Smart City Utilities~\thanks{Pre-print of article to appear in \emph{Software: Practice and Experience}, Wiley, 2018}} 

\author{Yogesh Simmhan~$^{\dagger,}$\thanks{Corresponding author. Email: \href{mailto:simmhan@iisc.ac.in}{mailto:simmhan@iisc.ac.in}}, Pushkara Ravindra~$^\dagger$, Shilpa Chaturvedi~$^\dagger$,\\ Malati Hegde~$^*$ and Rashmi Ballamajalu~$^*$\\
~\\
\em $^\dagger$~Department of Computational and Data Sciences,\\
\em $^*$~Department of Electronics and Communications Engineering,\\
\em Indian Institute of Science, Bangalore, India}

\date{}
\maketitle

\begin{abstract}
The Internet of Things (IoT) is emerging as the next big wave of digital presence for billions of devices on the Internet. Smart Cities are practical manifestation of IoT, with the goal of efficient, reliable and safe delivery of city utilities like water, power and transport to residents, through their intelligent management. A data-driven \emph{IoT Software Platform} is essential for realizing manageable and sustainable Smart Utilities, and for novel applications to be developed upon them. Here, we propose such a service-oriented software architecture to address two key operational activities in a Smart Utility -- \emph{the IoT fabric for resource management}, and \emph{the data and application platform for decision making}. Our design uses open web standards and evolving network protocols, Cloud and edge resources, and streaming Big Data platforms. \modc{We motivate our design requirements using the smart water management domain; some of these requirements are unique to developing nations.} We also validate the architecture within a campus-scale IoT testbed at the Indian Institute of Science (IISc), Bangalore, and present our experiences. Our architecture is scalable to a township or city, while also generalizable to other Smart Utility domains. Our experiences serves as a template for other similar efforts, particularly in emerging markets,
and highlights the gaps and opportunities for a data-driven IoT Software architecture for smart cities.
\end{abstract}

\section{Introduction}
\label{sec:intro}

The rapid emergence and deployment of Internet of Things (IoT) is causing millions of devices and sensors to come online as part of public and private networks~\cite{iot-scale}. This marks a convergence of cheap sensing hardware, pervasive wireless and wired communication networks, and the democratization of computing capacity through Clouds. It also reflects the growing need to leverage automation to enhance the efficiency of public systems and quality of life for society. While consumer devices such as smart-phones and fitness bands highlight the ubiquity of IoT in a digitally immersed lifestyle, of equal (arguably, greater) importance is the role of IoT in managing infrastructure such as city utilities and industrial manufacturing~\cite{simmhan:cise:2013,iiot}. Smart Cities and Industrial IoT deploy sensing and actuation capabilities as part of physical systems such as power and water grids, road networks, manufacturing equipment, etc. This enables the use of data-driven approaches to efficiently and reliably manage the operations of such vital \emph{Cyber Physical Systems (CPS)}~\cite{spe/JaraGB15,ibm,cps}. 

As the number of IoT devices soon reach the billions, it is essential to have a distributed software architecture that facilitates the sustainable \emph{management} of these physical devices and communication networks, and \emph{access} to their data streams and controls for developing innovative IoT applications. Three synergistic concepts come together to enable this. 
\emph{Service-Oriented Architectures (SOA)}~\cite{soa,Bouguettaya:2017:SCM} offer standard mechanisms and protocols for discovery, addressing, access control, invocation and composition of services that are available on the World Wide Web (WWW), by leveraging and extending web-based protocols such as HTTP and open representation models like XML~\cite{soa-ws}.  
\emph{Cloud computing} is a manifestation of this paradigm where infrastructure, platform and software resources are available ``as a service'' (IaaS, Paas and SaaS), often served from geographically distributed data centers world-wide. These offer economies of scale and enable access to elastic resources using a pay-as-you-go model~\cite{botta2016integration}. Such commodity clusters on the Cloud have also enabled the growth of \emph{Big Data platforms} that allow data-driven applications to be composed and scaled on tens or hundreds of Virtual Machines (VMs), and deal both with data volume and velocity, among other dimensions~\cite{vilajosana2013bootstrapping}.

Unlike traditional enterprise or scientific applications, however, the IoT domain is distinct in the way these technologies converge to support emerging applications. 
(1) IoT integrates hardware, communication, software and analytics, and links the physical and the digital world. Hence \emph{infrastructure management}, including of Cloud, Fog and Edge devices, is an intrinsic part of the software architecture~\cite{prateeksha:icfec:2017}. (2) These devices and services may not always be on the WWW, and instead connect within private networks or the public Internet (not WWW). Hence, \emph{network heterogeneity} is also a concern. (3) The communication connectivity and indeed even their hardware availability \emph{may not be reliable}, with transient network and hardware faults being common in wide-area deployments. (4) The \emph{scale} of the IoT infrastructure services (and micro-services) is likely to be orders of magnitude more than traditional business and eCommerce services, eventually reaching billions. (5) Lastly, the \emph{potential applications} that will be built on top of IoT is not yet well-defined and the scope of innovation is vast -- provided that the software architecture is open and extensible.

These necessitate a software architecture that encompasses a \emph{management fabric} and a \emph{data-driven middleware} that can leverage SOA, Clouds and Big Data platforms in a meaningful manner to support the needs of IoT applications. 
One can envision convergence onto a \emph{core set of interoperable, open standards} -- an approach that contributed to the success of the WWW using HTTP, HTML, URL, etc. specifications from IETF, or they may fragment into vertical silos pushed by \emph{proprietary consortia}, such as seen in public Clouds from Amazon AWS and Microsoft Azure (who themselves are evolving for IoT). Both can prove to be successful, but we argue the need for the former. \modc{The few major public Cloud providers have large customer bases. Hence, custom APIs that they offer based on web standards will have a captive market.} On the other hand, IoT will need to leverage open web and Internet standards, both existing and emerging, to allow interoperability and reuse of existing tools and software stacks. This is particularly of concern to developing countries with mega-cities that are transitioning to Smart Cities. Such an open-approach will also catalyze the development of novel applications for consuming the data and application services exposed by the city utility. 

\textbf{Contributions.} In this article, we propose a service-oriented and data-driven software architecture for Smart City utilities. This is motivated by representative applications for smart water management and validated for managing the infrastructure and applications within a Smart Campus IoT testbed at the Indian Institute of Science (IISc), Bangalore. We make the following specific contributions. (1) We characterize the requirements of an IoT fabric and application middleware to support innovative Smart City applications. (2) We develop a service-oriented software architecture, based on open protocols, standards and software, to meet these requirements while leveraging Cloud and Big Data platforms. This includes a novel bootstrapping mechanism to on-board new devices, and support for streaming synchronous and batch asynchronous analytics. (3) We integrate these technology blocks together within a real IoT field deployment at the IISc campus testbed, that spans sensing, communication, data integrating and analytics, to validate our design.

\textbf{Organization.}The rest of the article is organized as follows. First, in \S~\ref{sec:bg}, we offer a background of the IISc Smart Campus project and highlight the unique requirements of Smart City deployments in emerging nations like India. In \S~\ref{sec:comm}--\ref{sec:analyze}, we discuss different aspects of our proposed scalable, data-centric, service-oriented software architecture for the Smart Campus. This includes sensing and communication (\S~\ref{sec:comm}); management fabric for the devices and the network (\S~\ref{sec:fabric}); data platforms for data acquisition (\S~\ref{sec:acquire}); and Cloud and edge-based analytics for decision-making (\S~\ref{sec:analyze}). 
We contrast our work against related efforts globally in \S~\ref{sec:related}, and finally offer our conclusions and discuss future directions in \S~\ref{sec:conclusion}.

\section{Background}
\label{sec:bg}
We present an overview of the Smart Campus project at IISc that is developing an IoT management fabric and application platform for smart utility management. We use this, as well as our prior experience with the Los Angeles Smart Grid project~\cite{simmhan:cise:2013}, to motivate the unique needs of a Smart City software architecture.

\begin{figure}[t]
  \centering
       \includegraphics[width=0.65\columnwidth]{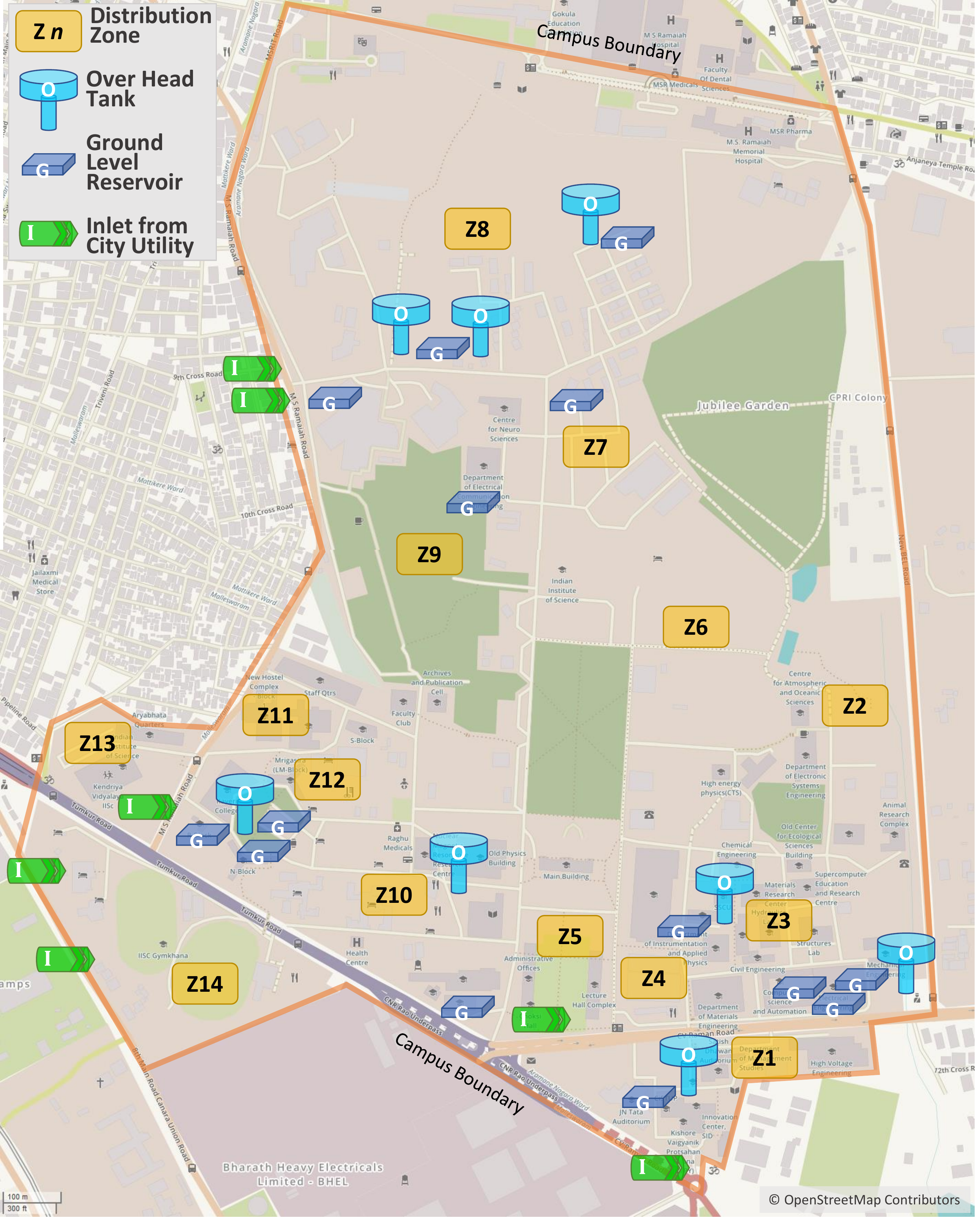}
  \caption{\modc{IISc Campus Map and Water Infrastructure}}
       \label{fig:campus}
\end{figure}
\subsection{IISc Smart Campus project}
The Government of India is undertaking a mission to upgrade $100$ cities into \emph{Smart Cities}~\footnote{Smart Cities Mission, Government of India, \url{http://smartcities.gov.in/}} over the next several years, at a cost of about USD~$14~billion$. While the exact characteristics of a ``Smart City'' are loosely defined, smart energy, water and waste management, urban mobility, and digital services for citizens are some of the thematic areas. Several township-scale and community-scale research and deployment projects have been initiated to understand the unique aspects of smart city management in a developing country like India, and the role of open technology in realizing this vision.

The \emph{Smart Campus project}~\footnote{IISc Smart Campus Project, \url{http://smartx.cds.iisc.ac.in}} at the Indian Institute of Science, the top graduate school in India, is one such effort to design, develop and validate a campus-wide \emph{IoT fabric}. This ``living laboratory'' will offer a platform to try novel IoT technologies and Smart City services, with a captive base of about $10,000$ students, faculty, staff and family who largely reside on campus. The gated campus spread across $1.8~km^2$ has over $50$ departments and centers, and about $100$ buildings which host offices, lecture halls, research labs, supercomputing facility, hostels, staff housing, restaurants, health center, grocery stores, and so on (Fig.~\ref{fig:campus}). This is representative of large communities and towns in India, and offers a unique real-world ecosystem to validate IoT technologies for Smart Cities.

The project aims to design, develop and deploy a \emph{reference IoT architecture} as a horizontal platform that can support various vertical utilities such as smart power, water and transportation management, with smart water management serving as the initial domain for validation of the fabric. In effect, the effort for this project is in sifting through and selecting the best-practices and standards in the public domain across various layers of the IoT stack, integrating them to work seamlessly, and validating them for one canonical domain at the campus scale. By its very nature, this limits \emph{de novo} blue-sky architectures that work in a lab setup but are infeasible, impractical, costly or do not scale. At the same time, the architecture also offers an open platform for research into sensing, networking, Cloud and Big Data platforms, and analytics.

IISc owns and manages the water distribution network within the campus, and in Bangalore, like other cities in India, water supply from the city utility is not $24\times7$ but rather periodic. As a result, there are under-ground reservoirs (ground-level reservoirs, GLR) to buffer the water from the city's inlets, large overhead tanks (OHT) where water is pumped up to from the GLRs, and rooftop tanks at the building-level where water is routed to from these OHTs using gravity. About $4$ city inlets, $13$ GLRs, $8$ OHTs, and over $50$ rooftop tanks form the campus water distribution network, and support an average daily consumption of $4~Million~cm^3$ of water. Fig.~\ref{fig:campus} shows these inlets, GLR and OHTs. The campus also consumes $10~MW$ of electricity, a tangible fraction of which goes to moving water between the storages.

The goal for \emph{smart water management} is to leverage the IoT stack to: (1) assure the quality of water that is distributed, (2) ensure the reliability of supply, (3) avoid wastage of water, (4) pro-actively and responsively maintain the water infrastructure, (5) reduce the costs of water and electricity used for pumping, and (6) engage consumers in water conservation. All of these will be achieved through \emph{domain-driven analytics} over the rich and real-time data that will be available on the water network from the IoT infrastructure.

\modc{The campus has $14$ water distribution zones that are grouped into $4$ logical regions for deploying and managing the network operations, as shown in Fig.~\ref{fig:campus}. Each region requires approximately $30$ wireless motes that transmit values sampled from sensors they are connected to. A gateway connects clusters of these nodes, and transmits the data to the Cloud through the campus network back-haul.}
A combination of water level and quality sensors, flowmeters, and smart power meters are used to sense the water network, with actuators for valve and pump controls planned. \modc{As we discuss later, the design of the \emph{ad hoc} wireless network is a key operational challenge.} At the same time, we need to ease the deployment, monitoring and management overheads of the IoT infrastructure.

These make for a unique validation environment for smart urban utility systems, with distinctive local challenges for observation, analytics and actuation, compared to developed nations. In contrast, a similar smart campus effort by the lead author at the University of Southern California (USC), Los Angeles, addressed challenges of demand-response optimization for Smart Grids~\cite{simmhan:cise:2013}. There, power was assured $24\times7$ by the Los Angeles Department of Water and Power (LA~DWP) but the goal was to change the campus energy use profile, on-demand, to reduce the load on the city power grid as more intermittent renewable sources are included within their energy portfolio. Also, the entire campus was instrumented using proprietary Smart Meters from Honeywell that worked off reliable wired LAN, could be centrally monitored and controlled using their custom software, had adequate bandwidth to push all data and analytics to the Cloud, and also carried a comparably high price tag for the solution \modc{-- such high-cost and proprietary solutions are impractical for emerging nations.}

\subsection{Desiderata}
\begin{figure}[t]
\centering
    \includegraphics[width=0.8\columnwidth]{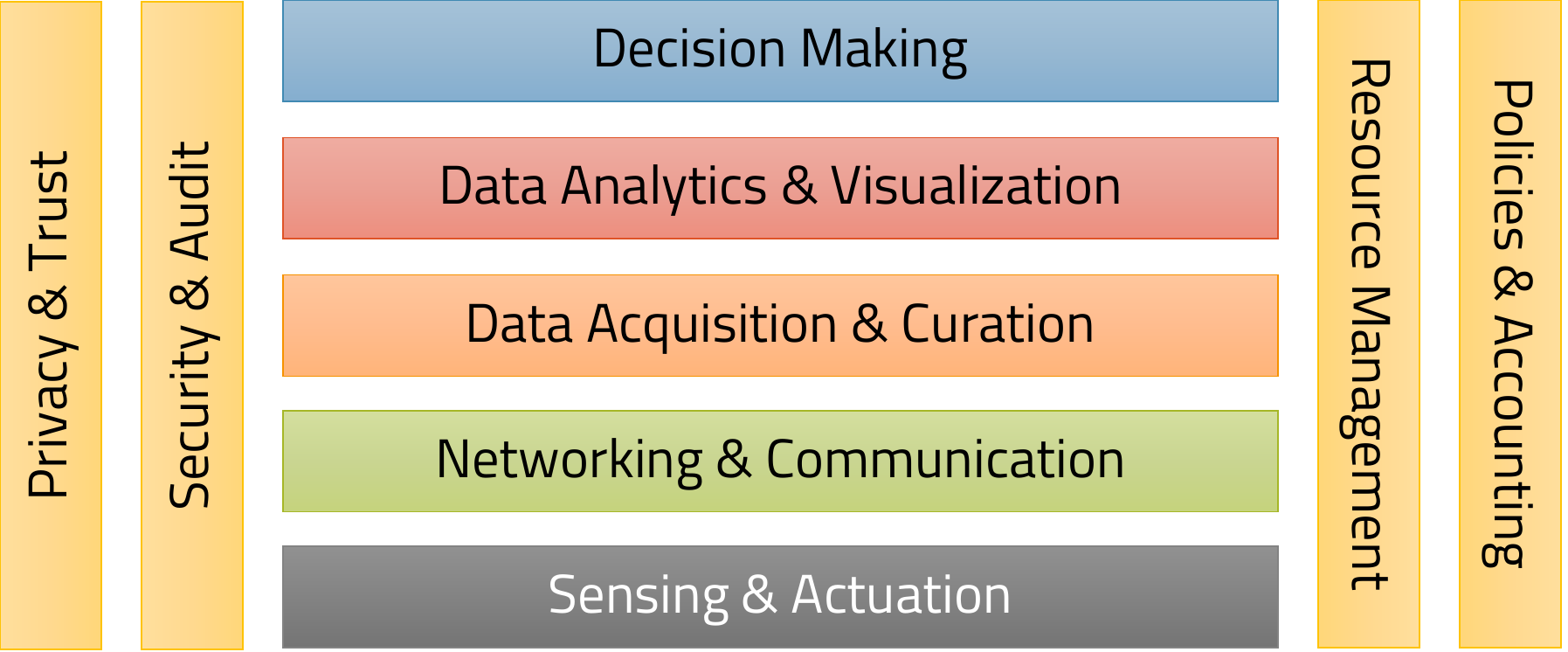}
  \caption{Functional IoT Architecture}
    \label{fig:func}
\end{figure}

Fig.~\ref{fig:func} illustrates the functional layers of an IoT software architecture, spanning sensors and communication, to data acquisition and analytics. We distinguish two parts to the IoT architecture. One, the \emph{IoT fabric} that manages the hardware and communication infrastructure and offers core resource management and networking. The other is the \emph{application platform} that acquires the data, and enables analytics and decision making that is fed-back to the infrastructure.

There are several guiding principles and high-level requirements for the software architecture~\cite{mishra:iotn:2015}.
\begin{enumerate}
\item \textbf{\modc{Scalablity.}}  Scalability of the architecture is paramount. The design should not have any intrinsic limitations or assumptions that prevent it from scaling city-wide even as the validation is for a township scale. The system should \emph{weakly scale} with the number of sensors, devices and motes that are part of the IoT infrastructure, the rate at which data is acquired, the number of domains and analytics, and the number of users of the system. 
  This recognizes the need to validate the design at small and medium scales to de-risk it before expanding to large scale urban environments, without fundamentally changing the model.

\item \textbf{\modc{Generalizability.}} The design should be generalizable enough to include additional utility domains such as smart power or waste management. While the sensors and analytics themselves can be domain dependent (or optionally shared across domains), the enabling fabric and platform layers must remain common across domains -- either conceptually or using different implementations or configurations.

\item \textbf{\modc{Modular Manageability.}} The architecture should allow new sensors, devices, data sources and applications to be included over time with limited overheads. The interface boundaries should be clearly defined to allow minimal configuration overheads. Support must be present for both static and transient devices, edge and Cloud resources, and for physical and crowd-sourced data collection and actuation.

\item \textbf{Reliability and Accessibility.} The architecture should monitor and ensure the health of the sensing, communication and computation layers, with autonomous actions where possible. Depending on the application domain, the QoS for data collection, decision making and enactment may be mission-critical or a best effort. Resource usage should be sensitive to current computing, networking and energy capacities. 

\item \textbf{Open Standards.} The architecture should use open protocols and standards, supported by standardization bodies and community efforts, as opposed to proprietary technologies or closed consortia. It will leverage existing open source Big Data platforms and tooling, and contribute to them to facilitate their growth.  It should balance the benefits of emerging IoT standards, and the reliability of mature ones, even if repurposed from other fields. It should be extensible and incorporate standards as they evolve.

\item \textbf{\modc{Cost Effectiveness.}} The design process will consider the costs for purchasing, developing, integrating, deploying and managing the architecture. These include hardware, software and service costs, as well as human capital to configure and maintain the IoT fabric, in the context of emerging nations (where human cost may be lower but technology costs higher). It should leverage commodity and open source technologies where possible. This recognizes that technology is a means to a sustainable end, rather than the end in itself. Designing such low-cost, innovative and sustainable technologies is locally termed as \emph{Jugaad}~\cite{jugaad}. 

\item \textbf{Security and Auditing.} The access to devices, data, analytics, and actuation services should be secured to prevent unauthorized access over the network, or even if the physical device is compromised. An audit trail must be established, and provenance must ensure data ownership and trust. Mechanisms for open data sharing and crowd-sourcing should be exposed, with a possible micro-payment model.
\end{enumerate}

\section{Sensing and Communication}
\label{sec:comm}
Sensing, actuation and communication are integral to the physical IoT fabric. The service-oriented software platform must be cognizant of their characteristics to allow for fabric management. Here, we discuss the capabilities and constraints of the edge and networking devices in the Smart Campus project for the water domain, which can be generalized to other utilities.

\subsection{Sensing and Actuation}
There are several types of physical sensors that are deployed for collecting real-time observations on the state of the water distribution network within the campus, and to perform demand-supply water balance studies. \emph{Flow meters} use electromagnetic induction to measure the volume of water flowing through the pipes in the distribution network, and \emph{pressure pads} observe the water pressure at various points in the network. These help us understand the flow of water through the major distribution lines across campus, and ensure sufficient pressure is available to deliver water. They are typically placed between the city inlet, the GLR and the OHT.
\emph{Smart power meters} at pumping stations let us know the energy usage for actively moving water between the various tanks, and can be correlated with the flow meters and pressure pads. In addition, \emph{water level meters} measure the depth of water in the OHT, GLR and the rooftop tanks continuously using ultrasonic signals to estimate the range from the top of these tanks to the water surface~\cite{verma2015towards}. By knowing the dimensions of the water tanks and when the pumps are operating, we can estimate the supply and the demand of water in individual buildings. These meters also record ambient temperature.

\begin{figure}[t]
 \centering
\begin{minipage}[c]{0.6\columnwidth}
\centering
  \subfloat[Water Quality Card]{
    \includegraphics[width=1.0\columnwidth]{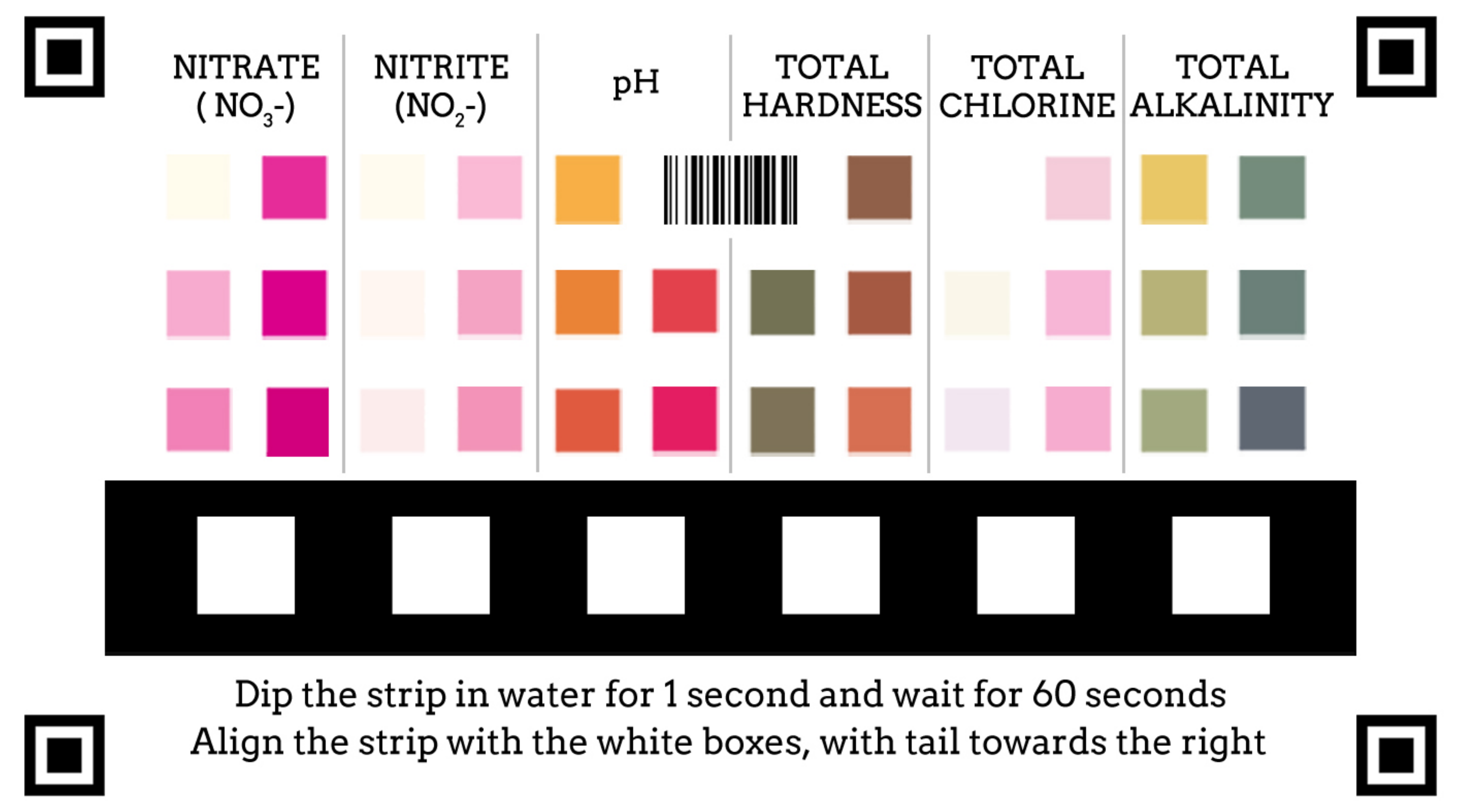}
    \label{fig:quality}
  }
\end{minipage}
~~~~
\begin{minipage}[c]{0.3\columnwidth}
\centering
  \subfloat[SmartWater Reporting App]{
    ~~~~\includegraphics[width=0.9\columnwidth]{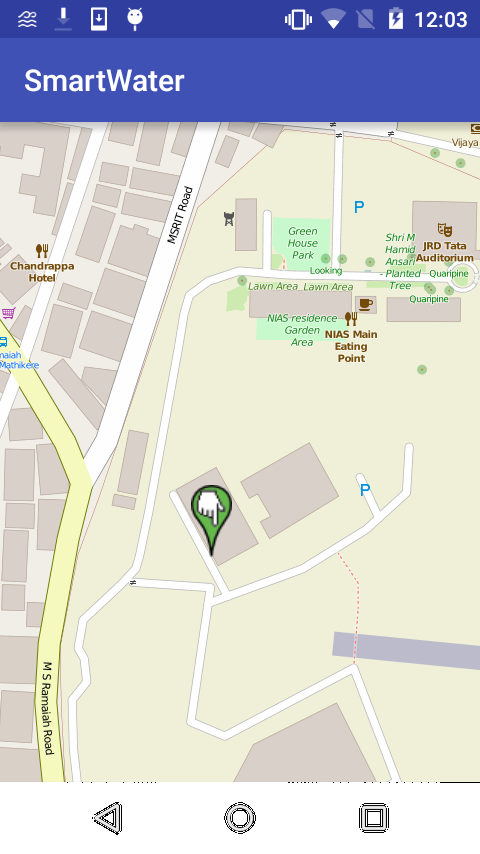}~~~~
    \label{fig:app}
  }
\end{minipage}
	\caption{Crowd-sourced data collection of water quality}
   \label{fig:crowd}
\end{figure}

The water level sensors can also serve as \emph{actuators} that control the pumps and the valves in the future. Physical actuators will automate the enactment of pumping and distribution decisions, in the absence of which, an SMS sent to a cell phone present with the pump operator can serve as a manual feedback control. Another form of actuation is to control the fabric itself. For example, the duty cycles of wireless motes and sampling rates for the various sensors and observation types can be controlled on the fly based on decision made by the management and analytics layers using information on the network, energy and computation resources, and the current application requirements.

Another important class of sensing and actuation within IoT is through \emph{crowd-sourcing} to supplement physical devices~\cite{cardone2013fostering}. Typically, crowd-sourcing can be used when the costs for deploying physical devices is high, or to engage the community through citizen science. Physical water quality sensors that can measure chemical properties are costly, and the number of potable water dispensers on campus is large. So we leverage the IISc residents in collecting quality measurements from dispensers that are distributed across buildings. Reagent strips available for $US\$0.25$ can be dipped in the water sample, placed against a water quality color card (Fig.~\ref{fig:quality}), and our Android smart phone app (Fig.~\ref{fig:app}) used to photograph and capture the color changes to the strip after normalizing for ambient light using the quality card~\cite{mit:little}. This reports water quality parameters such as nitrates, chlorine, hardness, pH, etc. The app can also be used to report maintenance issues such as water leakage and drips, water overflow or underflow in buildings, etc. Such participatory sensing engages the campus users in their own health, and instills a community value.

\subsection{Networking and Communication}
\label{sec:nw}

\begin{figure}[t]
  \centering
       \includegraphics[width=0.85\columnwidth]{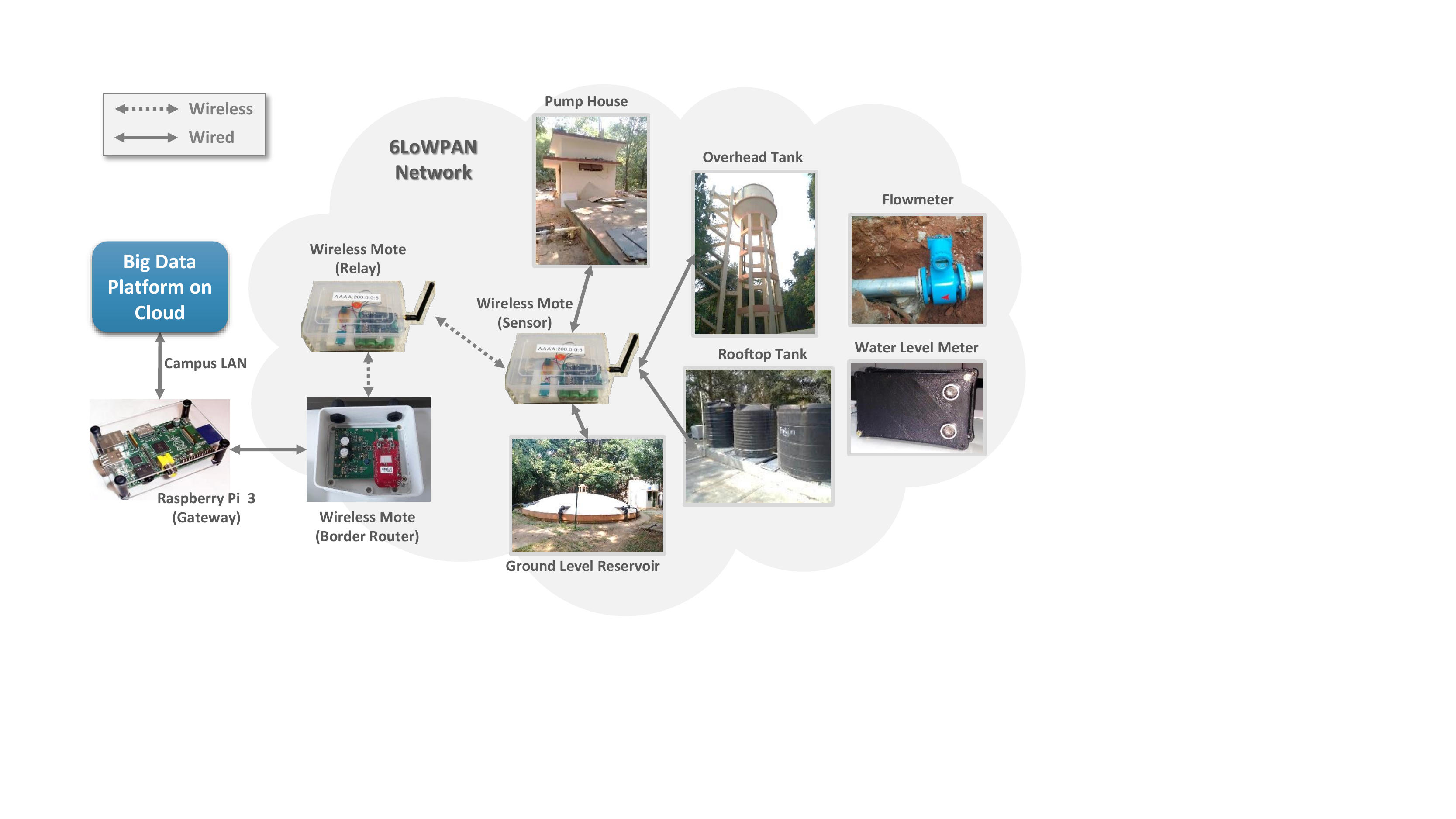}
       \caption{Wireless Sensor Network Deployment at IISc}
    \label{fig:nw-arch:wsn}
\end{figure}
\subsubsection{Network Protocols and Infrastructure} Communication networks are required to evacuate data from the sensors to the data platform or to trigger the actuators based on control decisions. Gateway devices and backend computing infrastructure hosting the platform, such as Cloud VMs, are on public or private infrastructure networks such as wired or wireless LAN. Accessing the sensors and edge devices becomes less challenging if such infrastructure networks are available within their vicinity, or if 2G/3G/4G connectivity can be made use of. However, field deployments may not be within range of LAN or WLAN, cellular connectivity may be costly, or devices that use these communication protocols may consume higher energy, which will be a constraint if they are powered by battery or solar renewable. 

As an alternative, \emph{ad hoc} and \emph{peer to peer (P2P)} network protocols
are popular for IoT deployments. There are multiple standards that can be leveraged here. \emph{Bluetooth Low Energy (BLE)} has gained popularity for Personal Area Networks (PAN) due to their ubiquity in smart phones. It is designed for P2P communication between proximate devices, such as smart phones and IoT beacons, within 10's of feet of each other, and supports 10's of kbytes/sec bandwidth. 

Alternatively, \emph{IEEE 802.15.4} specifies the physical (PHY) and media access control (MAC) protocol layers for PANs~\cite{802.15.4}. It operates in the unlicensed Industrial, Scientific and Medical (ISM) radio bands, typically $2.4~GHz$, and forms the basis for \emph{ZigBee}. It has been extended specifically for IoT usage as well. \emph{IEEE 802.15.4g} was proposed for P2P communications and for smart utility networks like gas, water and power metering. The Thread Group, including consortium members Samsung, Google Nest, Qualcomm and ARM, also use this standard for an IPv6-addressable \emph{Thread protocol} for smart home automation. 

More broadly, IETF's IPv6 over Low power Wireless Personal Area Networks \emph{(6LoWPAN)} extends IPv6 support for IEEE 802.15.4 on low power devices~\cite{6lowpan}. A single IPv6 packet has a Maximum Transmission Unit (MTU) of $1280~bytes$ which fits in traditional Ethernet links having an MTU of $1500~bytes$. But IEEE 802.15.4 only has an MTU of $127~bytes$, and 6LoWPAN acts as an adaptation layer to allow IPv6 packets to be fragmented and reassembled at this data link layer. It also enables IPv6 link-local auto-addressing and provides datagram compression.

The range and bandwidth of wireless networks depend on the
transmission power,
size of antenna, 
and the terrain. Typically, two of the three dimensions -- high bandwidth, low power, and long range -- are achievable. 
PANs choose a lower range in favor of higher speed and lower power. E.g., ZigBee with $2.4~GHz$ offers a range of $10-100~meters$, line of sight, and a bandwidth of $\approx30~kbytes/sec$. Using the \emph{sub-GHz} spectrum offers a longer range for Wide Area Networks (WAN), due to low attenuation of the low-frequency waves, but also a lower speed of $\approx5~kbytes/sec$. While IEEE 802.15.4g supports this frequency, \emph{LoRaWAN} technology has been developed specifically for such long ranges of a kilometer using, say, $868~MHz$ sub-GHz radio in the IoT context~\cite{lora}. LoRa uses a star-of-stars topology, and is well suited for applications with low data-rate of $0.25\sim5~kbytes/sec$, 
but the current implementation lacks support for the IP stack and uses a proprietary chipset.
\begin{figure}[t]
\centering
       \includegraphics[width=0.4\columnwidth]{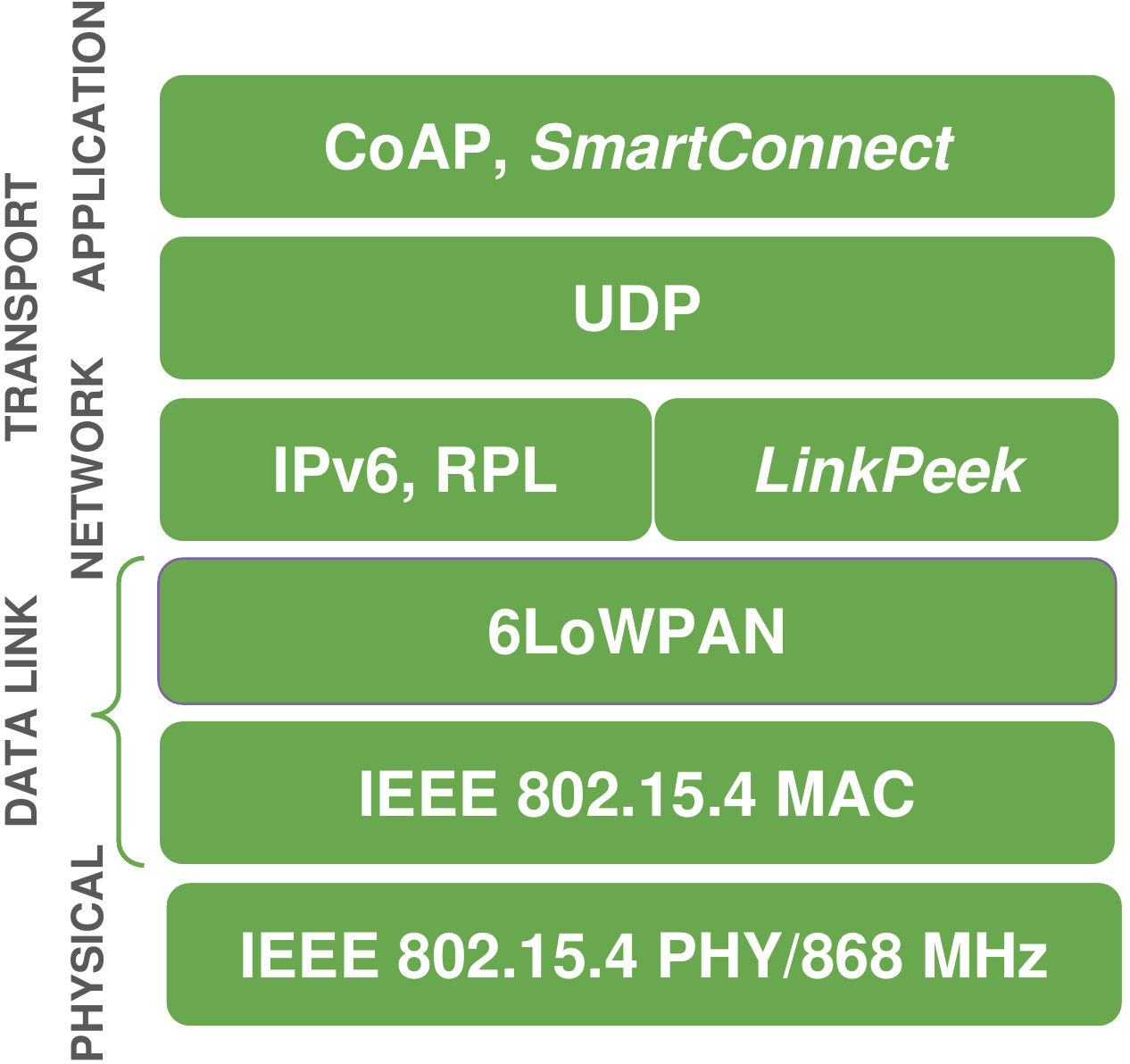}
       \caption{Network Protocol Stack of the IoT Fabric}
    \label{fig:nw-arch:stack}
\end{figure}

For the Smart Campus network fabric, the buildings have W/LAN access, and devices within WiFi range can use the backbone network. However, many of the OHT, GLR and pump houses are not in WiFi range. Hence, we deploy an \emph{ad hoc} 6LoWPAN Wireless Sensor Network (WSN) for such field devices (Fig.~\ref{fig:nw-arch:wsn} and~\ref{fig:nw-arch:stack}). We use Zolertia's \emph{RE-Mote}~\cite{remote} wireless hardware platform which has a dual-radio of a $2.4~GHz$ IEEE 802.15.4 and a sub-GHz $868/915~MHz$ RF transceiver.  It runs an 
ARM Cortex-M3 CPU 
at $32~MHz$ clock speed, with $512~KB$ of programmable flash and $32~KB$ of RAM. 
These motes connect to the sensors to acquire data and pass control signals, act as WSN relays, or are the border router connected to the gateway device. 

A \emph{Raspberry Pi 3} serves as the gateway that connects to the border router through a USB interface. Besides connecting the WSN to the campus backbone network, the Pi also acts as a proxy between the IPv6 WSN and the IPv4 campus network using the \texttt{tunslip} utility. 
Thus, all motes are IP addressable, with end-to-end IP-based connectivity across the campus. 
The use of such diverse network protocols coordinated through a gateway is generalized as an \emph{Area Sensor Network (ASN)} that serve as a bridging layer for composable regions of sensor networks that can scale to a city in a federated manner~\cite{molina:csmc:2014}.

A novel use of crowd-sourcing uses people as \emph{data sherpas} when sensors require many WSN hops to reach a building W/LAN but where human footfall is high~\cite{mishra:iotn:2015}. Here, data from the sensor is broadcast using a BLE beacon, which is picked up by the Smart Campus app on users' phones and pushed to our data platform through 3G/WiFi.
Lastly, small scale experiments using LoRaWAN is also being investigated. While they may be adequate for periodic water level or flow meter data, their bandwidth will limit the reuse of the network fabric for other data-heavy IoT domains.

\subsubsection{Network Deployment Design}

The WSN need to be designed and deployed across regions of the campus to ensure robust quality of service (QoS), and avoid data loss due to packet collisions and scattering of waves by dense buildings.
\emph{SmartConnect}~\cite{smartconnect} is an in-house tool for designing IEEE 802.15.4 networks. When given the sensor locations, their expected data traffic, the required QoS, and possible locations for relays, it identifies the lowest-cost relay placement with a given path redundancy in the multi-hop WSN.
SmartConnect uses two field measurements for pairwise placement of the motes at each candidate relay location: (1) the minimum \emph{Received Signal Strength Indicator (RSSI)} for which the \emph{Packet Error Rate (PER)} is consistently $\le 2\%$, and (2) the maximum radio reception distance, $R_{max}$, for which the packet delivery rate is $\ge 95\%$. 

\begin{figure}[t]
	\centering
   \subfloat[PER vs. RSSI]{
  	\includegraphics[width=0.45\textwidth]{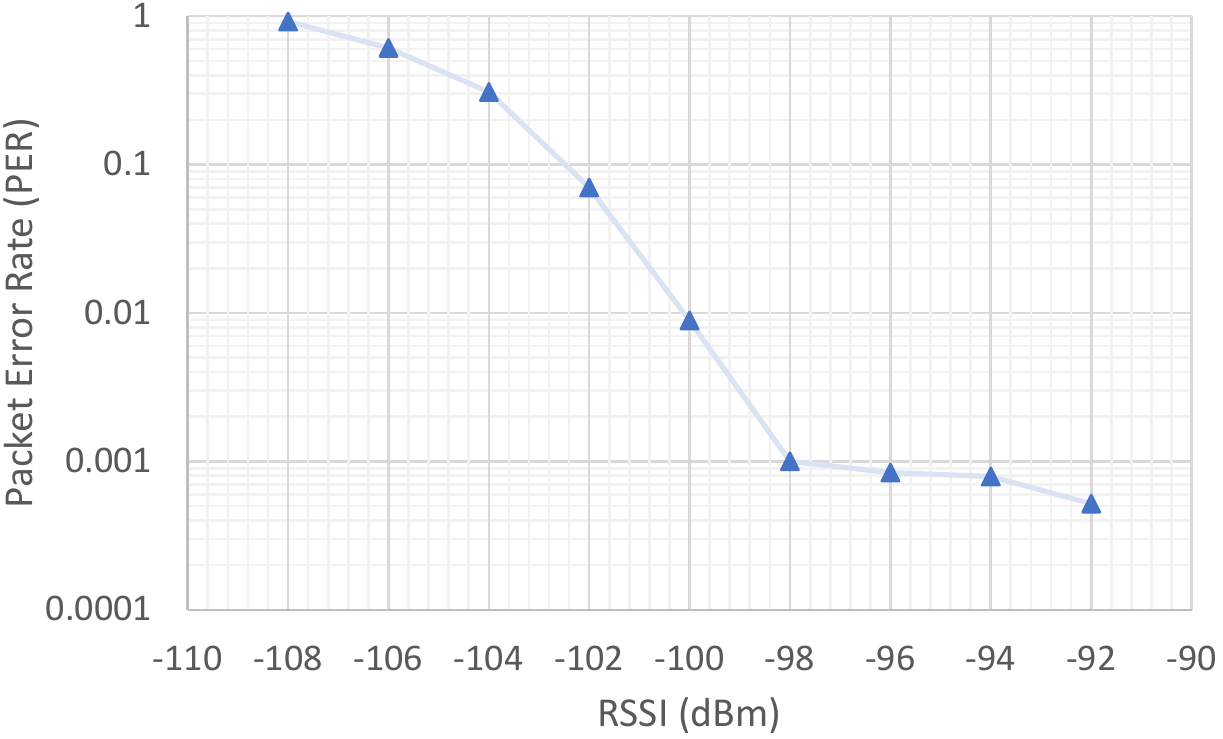}
  	\label{fig:nw:stats:per}
  }
  \subfloat[Fraction of links $P_{bad}$ with $\text{\emph{outage probability}}>P_{out}$ at different distance ranges.]{
  	\includegraphics[width=0.50\textwidth]{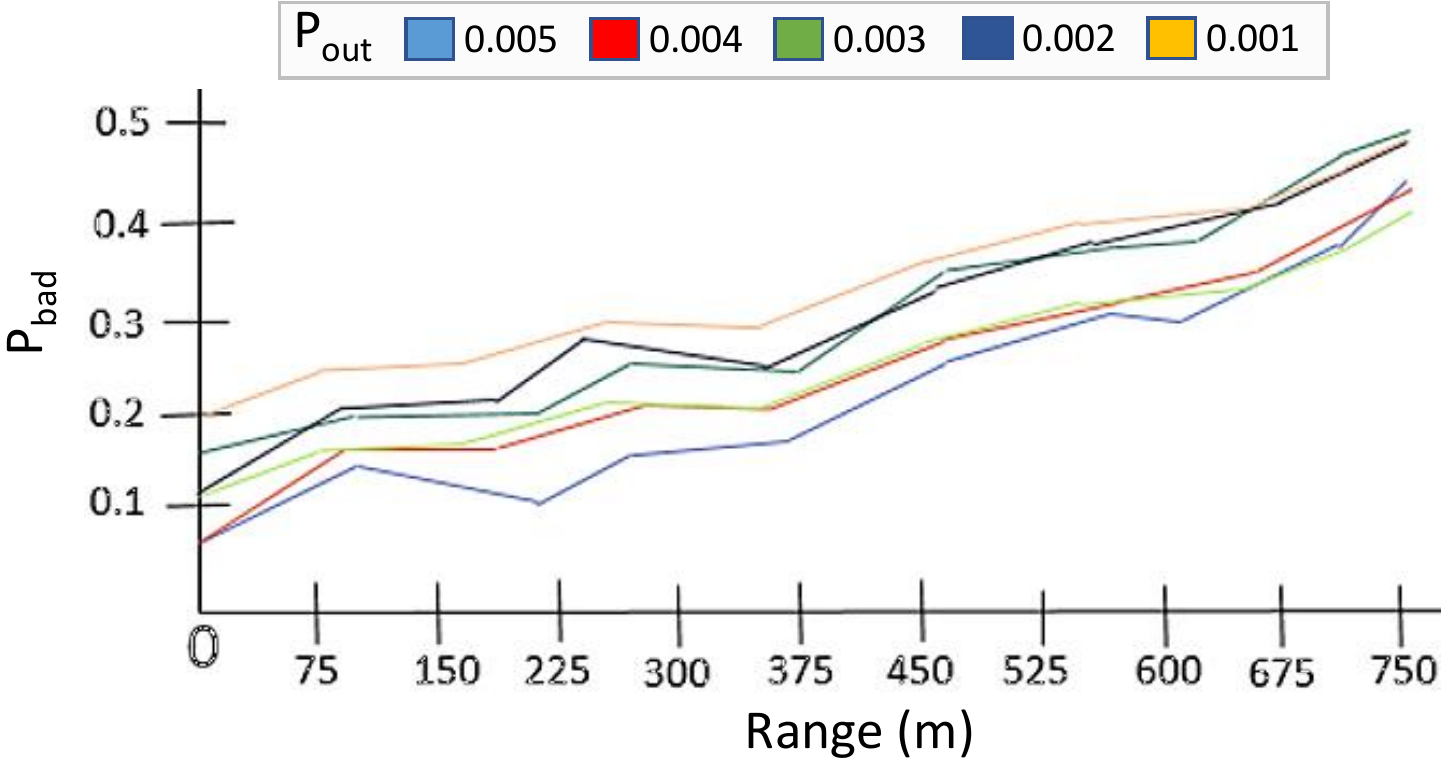}
  	\label{fig:nw:stats:range}
  }\\
\subfloat[\modc{Heatmap of RSSI Range at different locations on campus}]{
	\includegraphics[width=0.55\columnwidth]{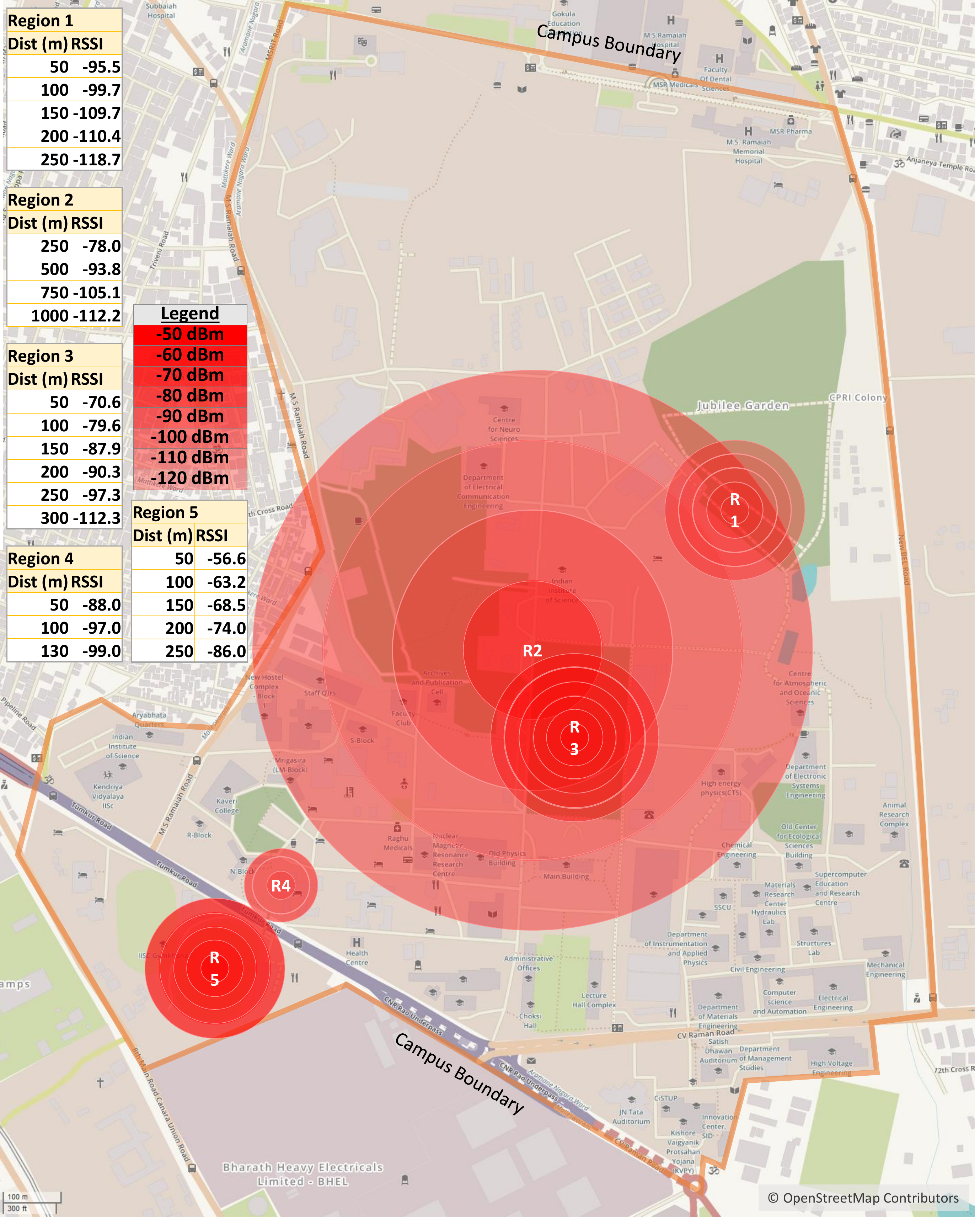}
        \label{fig:nw:heatmap}
}
\caption{Network characteristics for RE-Mote on the field.}
\label{fig:nw:stats}
\end{figure}

\modc{Fig.~\ref{fig:nw:stats} captures the results of several experiments with the Sub-GHz WSN deployed at different regions of campus to plan the deployment. Fig.~\ref{fig:nw:stats:per} shows the result of conducting wired back-to-back testing of motes to determine the optimal operation characteristics under ideal conditions. This offers a best-case baseline on the PER as we increase the signal strength, i.e., when inter-mote distance is not a concern. After calibrating the devices with the minimum RSSI, controlled experiments were conducted to obtain the practical operating distance range between motes for the required QoS as shown in Fig.~\ref{fig:nw:stats:range}. Here, $P_{out}$ indicates the upper bound of PER while $P_{bad}$ is the probability of a link having a PER worse than $P_{out}$, as the link distance is varied. Based on this, a minimum RSSI of $-97~dBm$ and a maximum range of $R_{max}=400~m$ were chosen for the field deployments. These were further validated on the field to capture the effect of topological characteristics on the network range, such as open spaces, buildings, tree cover, etc.~\cite{rathod:2015}. Fig.~\ref{fig:nw:heatmap} shows the heatmap of the signal strength and ranges. Here, R3 is in a wooded area and R4 is near dormitory buildings, and both show higher signal attenuation. R5 is measured near the recreational center with open spaces, and has a higher signal strength.}

Based on these experiments, for a QoS delay of $200~msec$, potential locations were suggested by SmartConnect for relay placement. These targeted experiments and analytical planning avoid having to actually deploy different permutations of the relays at every possible field location to determine the optimal placement for a reliable WSN. 

Once the motes are deployed, we implement the \emph{Routing Protocol for Low power lossy networks (RPL)}~\cite{rpl} for the formation and maintenance of the WSN (Fig.~\ref{fig:nw-arch:stack}). RPL maintains a Destination Oriented Directed Acyclic Graph (DODAG) among the motes, with every node having one or more multi-hop path(s) to the root, which is the border router. This supports multipoint-to-point (MP2P), point-to-multipoint (P2MP), and point-to-point (P2P) communication patterns. Packets traversing through such a multi-hop Low-power and Lossy Network (LLN) may get lost in transit due to various link outages at intermediate relay nodes. To ensure high Packet Delivery Ratio (PDR) in the LLNs running RPL, we include a lightweight functionality, \emph{LinkPeek}~\cite{linkpeek}, to the network layer's packet forwarding task. Here, the forwarding node iteratively retransmits the packet to its next best parent in the same DODAG whenever a preset MAC layer retransmission count for the current best parent is exceeded.

\section{IoT Fabric Management}
\label{sec:fabric}

\begin{figure}[t]
\centering
	\includegraphics[width=0.75\columnwidth]{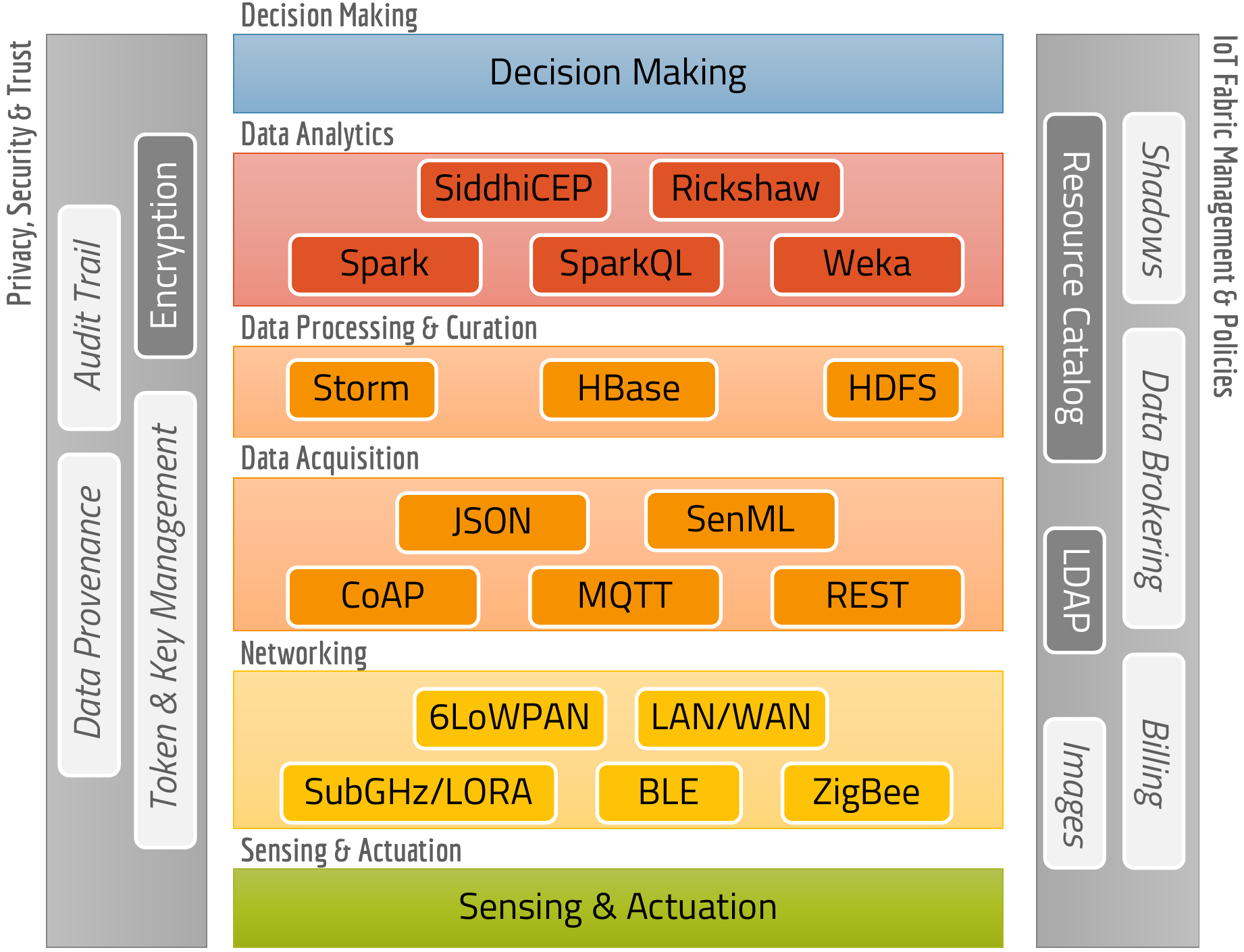}
	\caption{Protocols and standards used in the IoT architecture}
   \label{fig:proto}
\end{figure}

A high level protocol stack for the entire software architecture is shown in Fig.~\ref{fig:proto}. In this, \emph{fabric management} deals with the health and life-cycle of devices present in the IoT deployment. The primary devices that require this management are the sensors, actuators, motes and gateway devices that are physically deployed in the field. The fabric also ensures that endpoints are available to manage the devices and to acquire data or send signals. Here, we describe the service-oriented fabric management architecture for the IISc Smart Campus. 

\subsection{Service Protocols for Lifecycle and Discovery}

IETF's \emph{Constrained RESTful Environments} working group (CoRE WG)~\cite{core} is developing standards for frameworks that manage \emph{resource-oriented} applications in constrained environments such as IoT. It is intended to align with existing web standards like REST and HTTP, as well as emerging IoT network standards for IPv6. This makes it well suited for designing a standards-compliant service-oriented IoT architecture, and we leverage several specifications from CoRE.

\begin{figure}[t]
\centering
       \includegraphics[width=0.75\columnwidth]{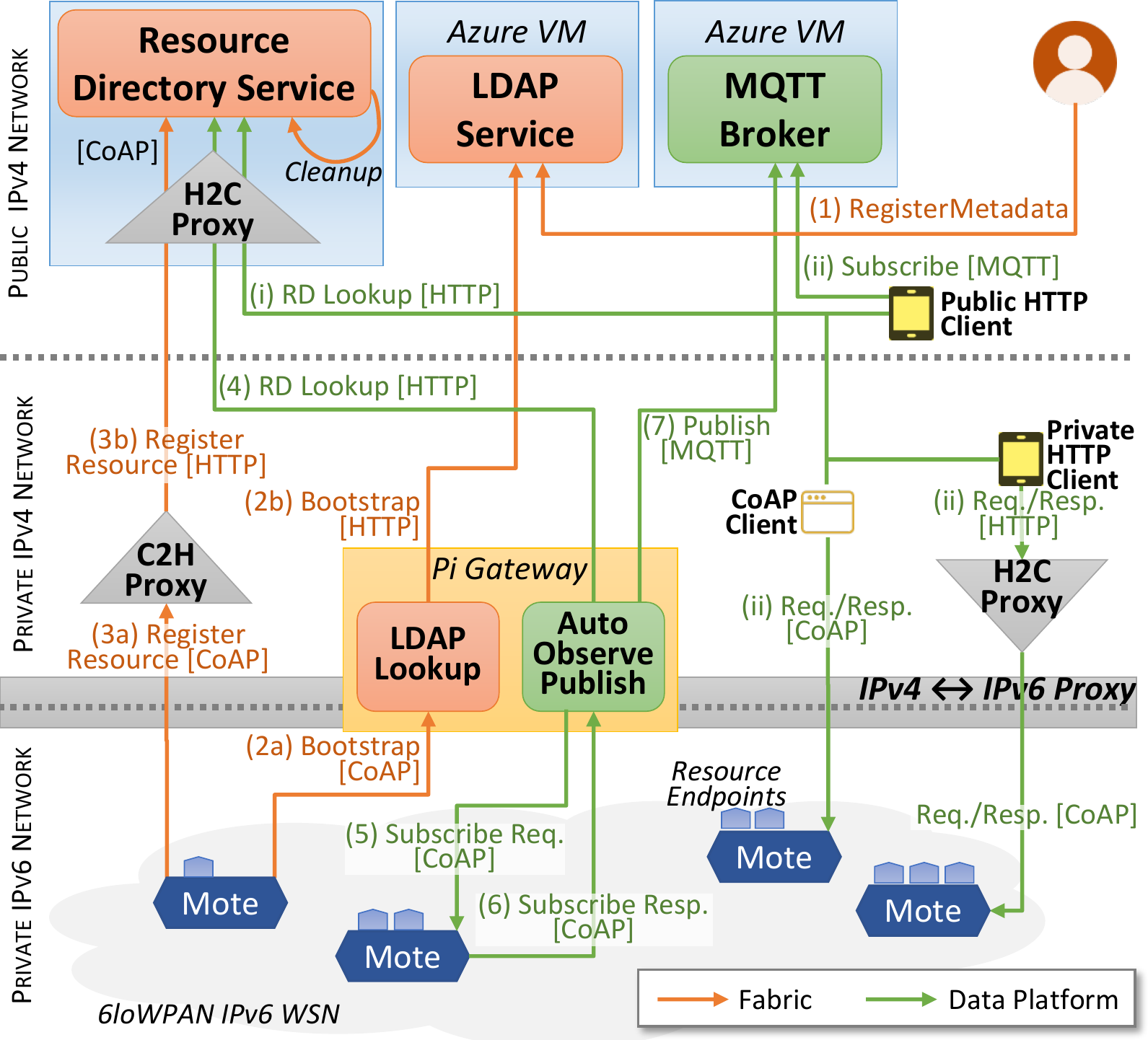}
       \caption{Interactions between architecture components for fabric management and data acquisition}
    \label{fig:fabric}
\end{figure}

Fig.~\ref{fig:fabric} shows an interaction diagram of various service components that enable fabric management (orange boxes and arrows). We adopt a \emph{stateful resource} model, similar to REST, for managing devices as services in the IoT deployment. These go beyond just the domain sensors and actuators, and also include network devices and gateways. Each device exposes one or more resources through a \emph{service endpoint}, each of which are either an \emph{observable} entity that can be sensed, or a \emph{controllable} entity that can be changed and the setup updated. E.g., a resource can represent domain observations, such as the water level or pump state, fabric telemetry, such as battery level of a mote, or a device setup state, such as sampling interval.

Two key services for the lifecycle management and discovery are the \emph{Light\-weight Directory Access Protocol (LDAP)}~\cite{ldap} and the \emph{CoRE Resource Directory (RD)}~\cite{rd}. Both of these are standards-compliant directory services, but play distinct roles in our design. LDAP is used to store \emph{static metadata} about various devices and resources that are, or can be, present in the IoT fabric. We use it as a \emph{bootstrapping} mechanism for devices to update their initial state during deployment. This reduces the overhead of deployment and configuration of the devices on the field, which may be done by a non-technical person, and instead have the device pull its configuration from the LDAP once it is online. RD, on the other hand, is responsible for maintaining the state and endpoint of resources that are currently active, and is used for \emph{dynamic discovery} of resources and interacting with them. RD supports frequent updates, and importantly, a \emph{lifetime} capability that automatically removes a service entry if it does not renew its lease within the interval specified when registering. This allows an eventually consistent set of active devices to be maintained in the RD, even if the devices do not cleanly de-register. Both these services are hosted on Cloud VMs to allow discovery by external clients, and sharing across private networks.

We adopt \emph{CoAP (Constrained Application Protocol)}~\cite{coap}, part of the CoRE specifications, as our service invocation protocol. CoAP is designed as the equivalent of REST over HTTP for constrained devices and well-suited for our 6LoWPAN network. CoAP has compact specification of service messages, uses UDP by default, has direct mappings to/from stateless HTTP protocol, and support Datagram TLS (DTLS) security. It has both request/response and observe/notify models of interaction, and offers differential reliability using confirmable/non-confirmable message types. We use CoAP as the default service protocol for all our devices on campus, including motes on the WSN and gateways like the Pi on the LAN. 

CoAP is an asynchronous protocol where requests and responses are sent as independent messages correlated by a token. It requires both the service and the client to be network addressable and accessible to each other -- this may not be possible for devices that are behind a firewall on a private network or data center. At the same time, while CoAP's use of UDP makes it light-weight within the private network, it can be lossy when operating over the public Internet. To address these two limitations, we switch from CoAP to traditional REST/HTTP over TCP/IP when interacting with services and clients on the public Internet from the campus LAN. Two proxy services present at the campus DMZ, \emph{CoAP to HTTP (C2H)} and \emph{HTTP to CoAP (H2C)}, enable this translation. Similarly, within the Cloud data center, we use an H2C proxy to switch back to CoAP in the private network to access the RD that is based on CoAP. 

While devices in the WSN are IP addressable and their CoAP service endpoints accessible by clients, they operate as an IPv6 network on 6LoWPAN. Hence, yet another proxy is present at the gateway device to translate between IPv4 used in the campus and the public network to IPv6 used within the WSN.
One of the advantages of leveraging emerging IoT standards from IETF and IEEE is that these protocol translations are well-defined, transparent and seamless.

These various services are shown in Fig.~\ref{fig:fabric}, and implemented using open source software, either used as is or extended by us to reflect recent evolutions of the specifications. We use the \emph{Eclipse Californium (Cf)} CoAP framework~\cite{californium} for the CoRE services such as Resource Directory, CoAP clients and services, and the C2H and H2C proxies on non-constrained devices that can run Java, such as the Pi and Cloud VMs. We also use the \emph{Erbium (Er)} CoAP service and client implementation for the ContikiOS running on the embedded mote platforms~\cite{ebrium}. The \emph{Eclipse Copper (Cu)} plugin for Firefox provides an interactive client to invoke CoAP services and browse the RD. \emph{Apache Directory Service} serves as our LDAP implementation.

\subsection{Device Bootstrapping and Discovery}

Each IoT device that comes online needs to determine its endpoint, the resources it hosts, and their metadata. Some are static to the device, while others depend on where the device's spatial placement. This device configuration during on-boarding has to be \emph{autonomic} to allow manual deployment of the last-mile field devices by non-technical personnel. We propose such an automated process for the bootstrapping using the LDAP for device initialization, and the RD for device discovery.

\begin{figure}[t]
	\centering
	\includegraphics[width=0.75\columnwidth]{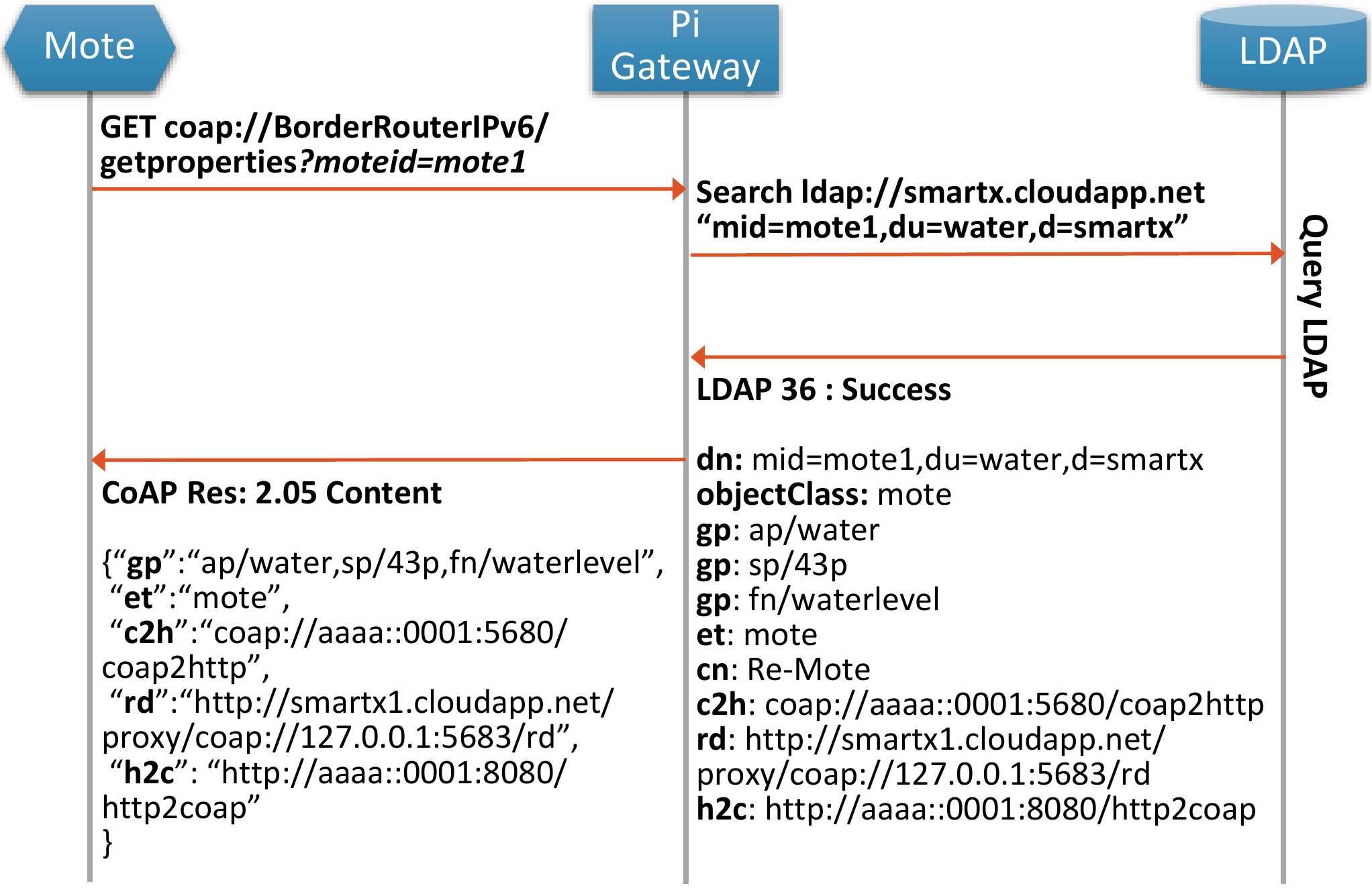}
	\caption{Sequence to bootstrap a device from LDAP.}
	\label{fig:bootstrap}
\end{figure}

Fig.~\ref{fig:bootstrap} shows the sequence diagram of messages for a device that comes online and connects to its gateway as part of the WSN -- a subset of these messages hold for devices not part of a WSN. Fig.~\ref{fig:fabric} shows the corresponding high level interactions. All IP-addressable devices in the deployment are considered as \emph{endpoints} that contain resources which are logically \emph{grouped}. These have to be auto-discovered based on minimal \emph{a priori} information. Each device is assigned, and will be aware of, just a globally unique UUID, and a ``well-known'' LDAP URL. Separately, an administrator registers the UUID and its metadata for all devices that will be deployed on the field in the LDAP directory information tree (DIT). The DIT is organized by domain, location, sensor type, etc. to allow group updates (\emph{(1)} in Fig.~\ref{fig:fabric}). 

\begin{figure}[t]
\centering
	\includegraphics[width=0.9\columnwidth]{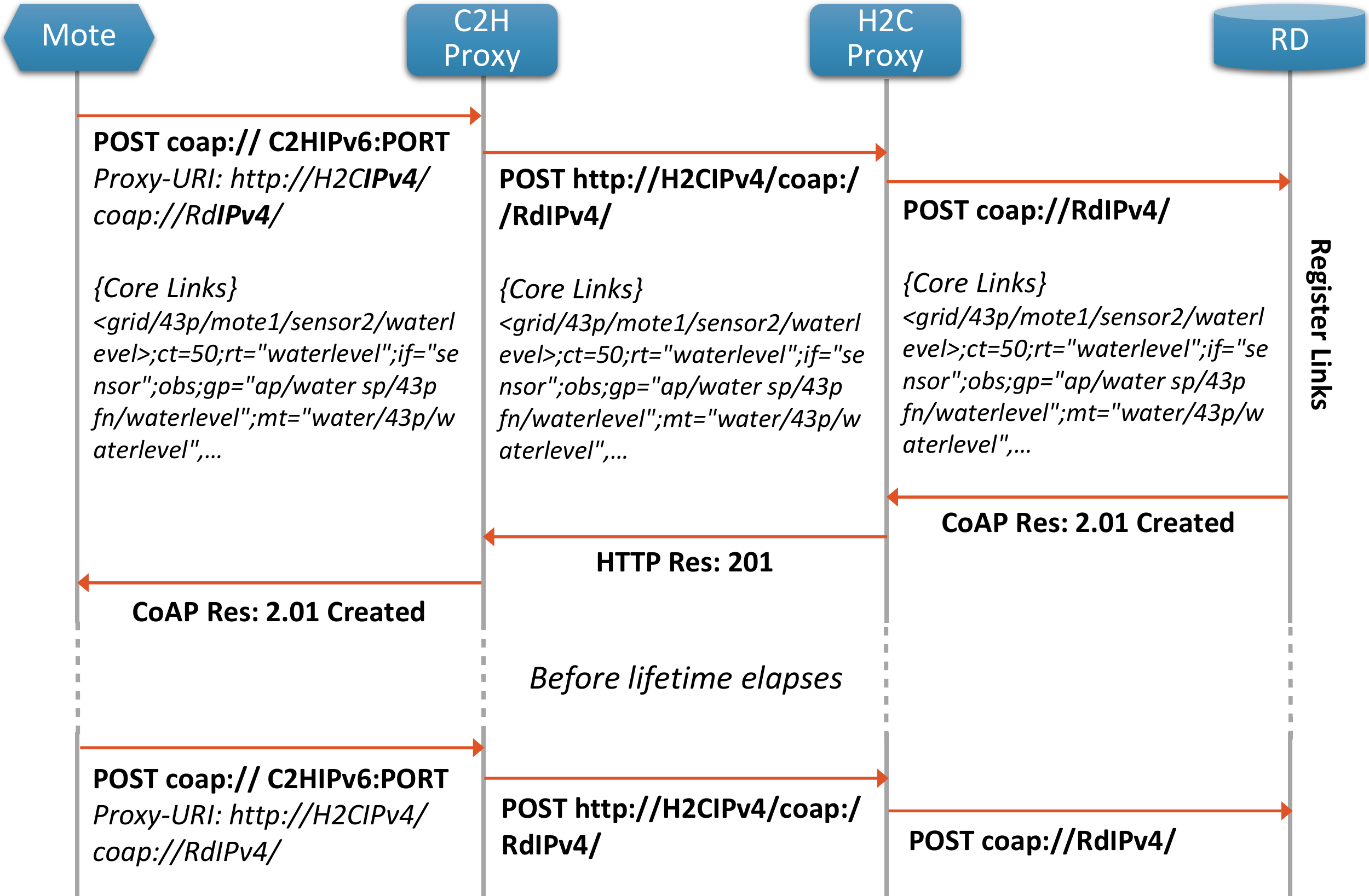}
	\caption{Sequence to register resources with RD \& renew lifetime.}
	\label{fig:rd}
\end{figure}

When a device connects to the campus IoT network, it does an HTTP query by UUID to the LDAP service for its metadata. Constrained motes, instead, perform a CoAP \texttt{GET} on an LDAP lookup service running on the gateway Pi, whose IP address matches the border gateway of the WSN. The Pi lookup service translates this to an LDAP HTTP query (Fig.~\ref{fig:bootstrap}; \emph{(2)} in Fig.~\ref{fig:fabric}). The response, optionally mapped from HTTP/LDIF to CoAP/JSON at the Pi, returns the entity type, its group(s), Distinguished Name (DN), spatial location, etc., and global URLs for the proxy services, RD, MQTT broker, etc. (Fig.~\ref{fig:bootstrap}). We use a well-defined rule to generate \emph{unique URI paths} for resources at this endpoint based on their metadata, which combines the spatial location, device and sensor type, and observation type, as shown below.

After a device is bootstrapped, it needs to register the resources available at its endpoint (ep) with the RD so that users or Machine-to-Machine (M2M) clients can discover their existence. The RD uses the \emph{CoRE link format}~\cite{link}, based on HTTP Web Linking standard, for this resource metadata. Each CoRE link contains the URI of the resource -- the optional endpoint hostname/IP:port, and the URI path -- along with the \emph{resource type (rt)}, the \emph{interface type (if)}, and the \emph{maximum size (sz)} of a \texttt{GET} response on this resource. Further, the RD also allows specifying the \emph{content type (ct)} such as JSON, the \emph{groups (gp)} the resource belongs to, and if the resource is \emph{observable (obs)}, i.e., can be subscribed to for notifications~\cite{observe}. Lastly, we use the extensibility of CoRE links to include an \emph{MQTT topic (mt)} parameter for observable resources which will publish their state changes to this topic at a publish-subscribe broker (\S~\ref{sec:mqtt}). 

Below is a sample CoRE link for an endpoint path \texttt{`grid/43p/mote1/sensor2/waterlevel'} with an observable \texttt{`waterlevel'} resource from \texttt{`sensor2'} that is attached to \texttt{`mote1'} placed at UTM grid location \texttt{`43p'} and returning JSON content type (\texttt{`ct=50'}).

{\textcolor{BLUE}{
\centering\texttt{<grid/43p/mote1/sensor2/waterlevel>;ct=50;rt="waterlevel";\\
if="sensor";obs;gp="ap/water~sp/43p~fn/waterlevel";\\
mt="water/43p/waterlevel"\\
}
}
}

\begin{figure}[t]
\centering
	\includegraphics[width=0.7\columnwidth]{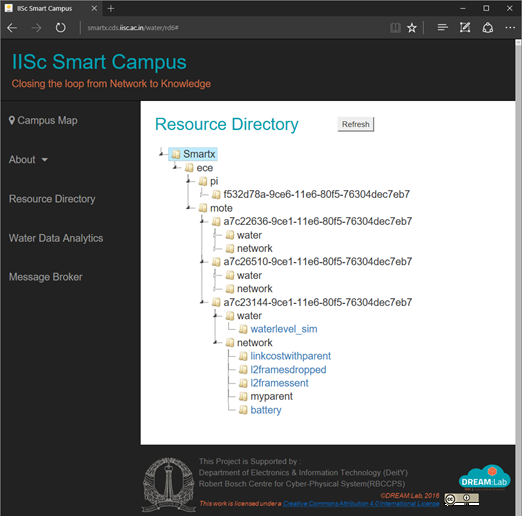}
	\caption{Portal displaying RD entries for the ECE building. 1 Pi and 3 mote endpoints each have multiple resources.}
	\label{fig:rdviz}
\end{figure}

Fig.~\ref{fig:rd} shows the sequence of operations for the device to register its resource(s) with the RD. Note the use of the C2H and H2C proxies to translate from CoAP within campus to HTTP on the public Internet, and back to CoAP within the VM hosting the RD. Registrations with the RD should also include a lifetime (\emph{lt}) for the entry in seconds, with the default being $24~hours$. If the resource does not renew this within this lifetime, the RD removes this entry and the resources are presumed to be unavailable. Clients can browse the RD (Fig.~\ref{fig:rdviz}), or query it using its CoAP or HTTP REST proxy API to discover resources of interest, and subsequently interact with the resource endpoint using CoAP.

\subsection{Monitoring and Control}
We make use of service endpoints to monitor the health and manage the configuration of devices such as motes and gateways as well. All motes expose CoAP resources to monitor their telemetry such as battery voltage, link cost with parent, and frames dropped, while gateway Pis report their CPU, memory and network usage statistics. These go beyond the liveliness that RD reports, and is in real-time. They help monitor the health of the network and device fabric, and take preventive or corrective actions, say, if a mote exhibits sustained packet drops or a Pi's memory usage becomes high. While some issues may require personnel on the field to fix things, others may be resolved remotely using control endpoints, such as restarting a mote or changing the sampling interval to reduce battery usage cost or packet drops. The analytics platforms, introduced later, that support the domain applications are also leveraged for such decision-making to optimize the IoT infrastructure.

\section{Data Acquisition and Storage}
\label{sec:acquire}

One of the characteristics of IoT applications is the need to acquire data about the system in real-time and make decisions. Given an operational IoT deployment and the ability to discover resources for observable and controllable devices, the next step is to acquire data about the utility infrastructure, and pre-process and persist them for downstream analytics. Data acquisition from $100-1000's$ of sensors has to happen at scale and with low latency, and from constrained devices. Once acquired, these streams of observations have to be transformed and validated at fast rates to ensure data quality. We make a design choice to integrate all observation streams in the Cloud to allow us to utilize scalable VMs and platform services, and collocate real-time data with historic data in the data-center on which analytics are performed. Next, we discuss our approach of using publish-subscribe mechanisms and fast data platforms for these needs.

\subsection{Asynchronous Access to Publish-Subscribe Observations}
The transient nature of sensor resources and the diverse applications and clients that may be interested in their observations means that using a synchronous request-response model to poll the resource state will not scale. Further, the rate at which the observations change may be infrequent for many sensors (e.g., the water level, or even battery level, gradually drains) and repetitive polling is inefficient. Rather, an asynchronous service invocation based on a subscription pattern is better suited. Here, the client registers interest in a resource, and is notified when its state changes.

We explore two mechanisms for such asynchronous observations of sensors, leveraging the native capabilities of CoAP and the scalable features of MQTT message brokers that are designed for IoT.

\subsubsection{CoAP's Observe Pattern}

CoAP services have an intrinsic ability to transmit data by subscription to clients interested in changes to the resource state~\cite{observe}. CoAP resources that indicate in their CoRE link format as being \emph{observable} allow this capability, and it complements the request-response model. Clients (\emph{observers}) can register interest in a resource (\emph{subject}), and the service then notifies the client of their updated state when it changes. The resource can also be parameterized to offer flexibility in terms of what constitutes a ``change'', say, by passing a query that observes changes to a moving average of the resource's state, or when a certain time goes by since the last update. The service maintains a list of observers and notifies them of their state change, but is designed to be eventually consistent rather than perfectly up to date. This ensures that the CoAP service is not frequently polled, making it amenable to the compute and network constrained environments like 6LoWPAN.

All our motes expose this capability for their fabric resources and the sensor resources \modc{that} they are connected to. This model, however, does have its limitations. It requires the service to maintain the list of observers, which can grow large and unmanageable for constrained devices. Further, this is a point-to-point model and each observer has to be individually invoked to send the notification, duplicating the overhead. Also, current open-source software support for \modc{CoAP} is limited to only resource state changes without any parameterization, though this is expected to change.

\subsubsection{MQTT Broker}
\label{sec:mqtt}
Publish-subscribe (or pub-sub)~\cite{pubsub} is a messaging pattern that is asynchronous, and \modc{uses a hub-and-spoke rather than point-to-point communication.} Here, the source of the message (\emph{publisher}) is not directly accessed by the message consumer(s) (\emph{subscriber(s)}). Instead, the messages are sent by the publisher(s) to an intermediate \emph{broker} service, which forwards a copy of the message to interested subscribers. The message routing may be based on topics (like a shared mailbox), or the type or content of the message.
The pub-sub pattern is highly scalable for IoT since the publishers and subscribers are agnostic to each other. This ensures loose coupling in the distributed environment while reducing their management overheads. Also, we drop from $m \times n$ messages set between $m$ publishers and $n$ subscribers to $m + n$ messages, avoiding duplication. The publishers and subscribers can also be on different private networks, and use the public broker for message exchange.

We use the \emph{Message Queue Telemetry Transport (MQTT)} ISO standard which was developed as a light-weight pub-sub protocol for IoT~\cite{mqtt}. Publishers can publish messages to a \emph{topic} in the broker, and subscribers can subscribe to one or more topics, including wildcards, to receive the messages. The topics have a hierarchical structure, allowing us to embed semantics into the topic names. Clients initiate the connection to the broker and keep it alive, allowing them to stay behind firewalls as long as the broker is accessible. The control payload is light-weight. The last published message to a topic may optionally be retained for future subscribers to access. It also supports a ``last will and testament'' message that is published to the \emph{will topic} if the client connection is killed, letting subscribers know of departing publishers. Three different delivery QoS (and costs) are supported -- at most once (best effort), at least once, and exactly once.

We use the \emph{Apache Apollo} MQTT broker implementation hosted in a VM in the Cloud as part of our IoT platform stack. It supports client authentication and TLS security. Topics are created for observable resources in the Smart Campus based on a production rule over the resource metadata, including the domain, spatial location, device and observation types, and the UUID for the device. This allows wild-card subscriptions, say, to all \texttt{waterlevel} messages or all messages from the \texttt{ECE} building. The MQTT topic is present in the CoRE link registered with the RD, allowing the discovery and subscription to these topics.

Non-constrained devices like the Pi gateways and devices on the public network, such as the Android App, publish their resource state changes and observations to the MQTT broker. For reasons we explain next, constrained devices do not \emph{directly} publish to the broker. We adopt IETF's \emph{Sensor Markup Language (SenML)} for publishing observations to topics~\cite{senml}. This offers a self-descriptive format for time-series observations, single and multiple data points, delta values, simple aggregations like sum, and built-in SI units. It also has well-defined serializations to JSON, CBOR, XML and EXI. Clients interested in the real-time sensor streams, such as our data acquisition platform, Smart Campus portal (Fig.~\ref{fig:streamviz}), and the Smart Phone app, subscribe to these topics and can visualize or process the SenML observations.

\subsubsection{Automated Observe and Publish from Gateway}
\begin{figure}[t]
	\centering
	\includegraphics[width=0.75\columnwidth]{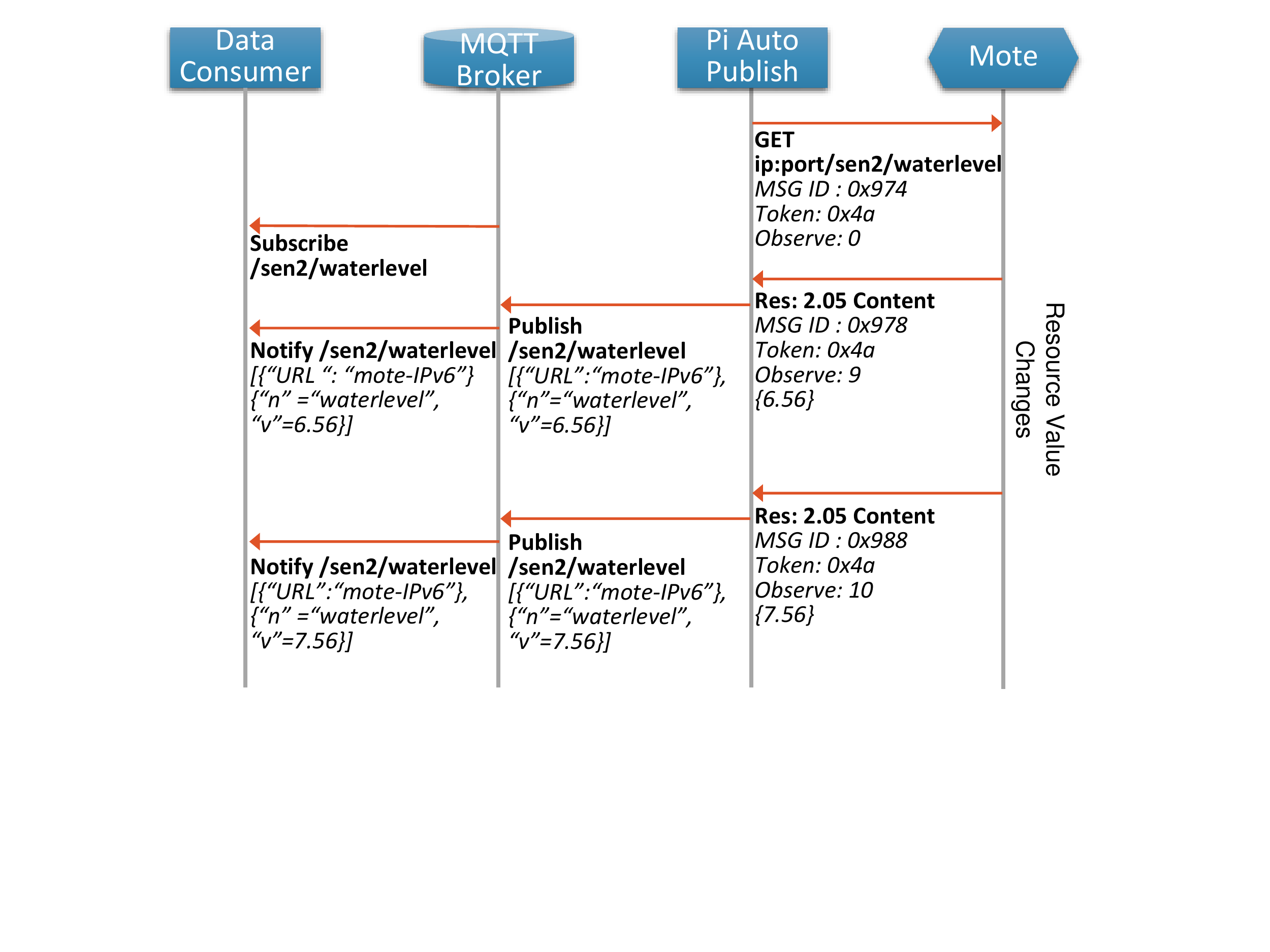}
	\caption{Sequence for data acquisition from sensors \modc{using AOP}. The gateway initiates a CoAP \texttt{observe} and auto-publishes SenML values to MQTT. Clients can subscribe to the MQTT topic.}
	\label{fig:observe}
\end{figure}

Publishing \modc{directly} to the MQTT broker is still heavyweight for our constrained devices and WSN for several reasons. One, is the overhead to initiate and keep the network connection open to the broker. Two, is the memory footprint for the MQTT client library on these embedded platforms, besides the CoAP service. Third, our choice to publish SenML causes a payload much larger than the native observations.

In order to offer the transparency of the pub-sub architecture while keeping with the limitations of the devices and WSN, we develop an \emph{Automated Observe and Publish (AOP)} service at the Pi gateway that couples the CoAP Observe capability with the MQTT publisher design. This is illustrated in Fig.~\ref{fig:fabric}, and the sequence of operations \modc{is} shown in Fig.~\ref{fig:observe}. This service on the Pi periodically queries the RD for new resources registered in the WSN group it belongs to (\emph{(4)} in Fig.~\ref{fig:fabric}; Fig.~\ref{fig:observe}). If discovered, the AOP service registers an \texttt{observe} request with the service endpoint for all new resources (\emph{(5)} in Fig.~\ref{fig:fabric}). When the endpoint notifies AOP of an updated resource state (\emph{(6)} in Fig.~\ref{fig:fabric}), AOP maps them to SenML/JSON and, as a data proxy, publishes them to the MQTT topic for that resource as listed in its CoRE link in the RD (\emph{(7)} in Fig.~\ref{fig:fabric}).

This design has the additional benefit of allowing clients that are interested in the observation to subscribe to the MQTT broker on the Cloud VM rather than the CoAP service on the constrained device. Consumers in the private network that are latency sensitive can always use the CoAP observe feature, or poll the service directly, and avoid the round trip time to the MQTT broker. \modc{E.g., Figs.~\ref{fig:perf:nw:lat} and~\ref{fig:perf:nw:bw} show the round trip latency and the bandwidth of pairs of Pi's within the campus backbone network, and between the Pi gateway devices on campus and the Azure VMs at Microsoft's Singapore Cloud data center. These violin plot distributions are sampled over a $24~hour$ period, and indicate the Edge-to-Edge and Edge-to-Cloud network profiles~\cite{ghosh:tcps:2017}. We see substantial latency benefits in subscribing to the event streams from within the campus network, which has a median value of $10~ms$ (green bar), compared to $153~ms$ when publishing to the Cloud. However, some regions of the campus have to go through multiple network switches and their latencies approach that of moving to the Cloud, as shown by the higher mean value (red bar). The bandwidth within campus is also $50\%$ faster and tighter, compared to between campus and Cloud. Our IoT middleware offers multiple means of accessing the observation streams to allow applications to choose the most appropriate one based on their presence in the network topology.}

\begin{figure}[t]
	\centering
	\includegraphics[width=0.85\columnwidth]{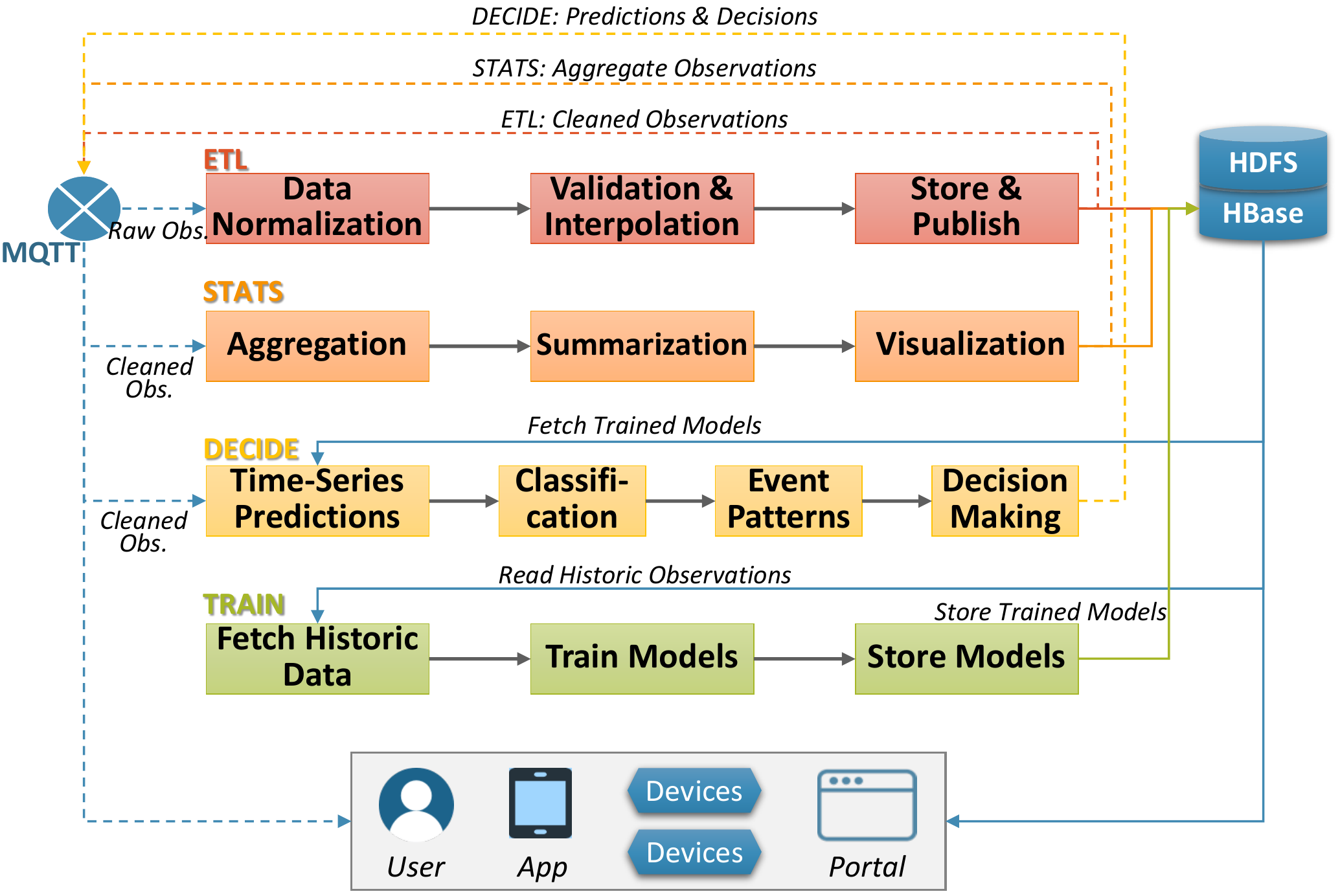}
	\caption{Interactions between streaming data acquisition and analytics dataflows in the data platform hosted on the Cloud}
	\label{fig:pipelines}
\end{figure}
\subsection{Fast Data Processing and Persistence}

Once data is published to the MQTT broker in the Cloud, there is a multitude of Big Data platforms that can be leveraged for processing the sensor streams in the Cloud data center. We take an approach similar to our earlier work~\cite{simmhan:cise:2013}, but with contemporary data platforms and updated domain logic relevant to the IISc Smart Campus.

Data published to MQTT needs to be subscribed to and persisted as otherwise these transient sensors streams are lost forever. At the same time, the data arriving from heterogeneous sensors have to be validated before they are used for analytics and decision making, such as turning off pumps or notifying users of a water quality issue. Hence, the cleaned observations should be available with limited delay. \emph{Distributed Stream Processing Systems (DSPS)} are Big Data platforms tailored for applications that need to process continuous data streams at high velocity within low latency on commodity cluster and Cloud VMs~\cite{dsps}. DSPS allow users to compose persistent applications as a dataflow graph, where task vertices have user logic, and edges stream messages between the tasks. 

\modc{There are several contemporary DSPS such as Apache Storm, Flink, Spark Streaming, Azure HDInsight, etc. We choose to use the \emph{Apache Storm} DSPS~\cite{storm} from Twitter due to its maturity and active open-source support, and its ability to compose a Directed Acyclic Graph (DAG) of modular user-defined tasks, rather than just higher order primitives.} Storm is used as our data acquisition platform for executing several streaming dataflow pipelines on sensor observations published to the MQTT broker (Fig.~\ref{fig:pipelines}). Two important and common classes of dataflows are \emph{Extract-Transform-Load (ETL)} and \emph{Statistical Summarization (STATS)}~\cite{shukla:riot:2017}. 

The ETL pipeline helps address data format changes and quality issues before storing the observations. The input to ETL is by subscribing to wild-card topics in the MQTT broker by sensor type, which allows all observation types supported by this pipeline to be acquired. Care is taken to cover all relevant topics so that no observation stream is lost; alternatively, it can query the RD to subscribe to specific topics in the broker, or use a special advertisement topic when new devices are on-boarded. The incoming messages may arrive from heterogeneous sources in different measurement units and formats, though SenML is preferred. Tasks like parsing, format and unit conversion help normalize these observations. There can also be missing or invalid values, say, due to network packet drop or sensor error. For example, we see the water level sensor report incorrect depths due to perturbation in the water surface or sunlight reflecting into the ultra-sonic detector. Range filters, smoothing and interpolation tasks perform such basic validation, quality checks and corrections. Lastly, the raw and validated data will need to be stored for future reference and batch analytics. We use \emph{Hadoop Distributed File System (HDFS)} to store the raw observations from MQTT and the \emph{HBase NoSQL database}~\cite{hbase} to store the cleaned time-series data for batch analytics.
The ETL dataflow also publishes the resulting cleaned sensor event stream to an MQTT topic which then can be subscribed to by downstream applications.

\begin{figure}[t]
\centering
    \includegraphics[width=1.0\columnwidth]{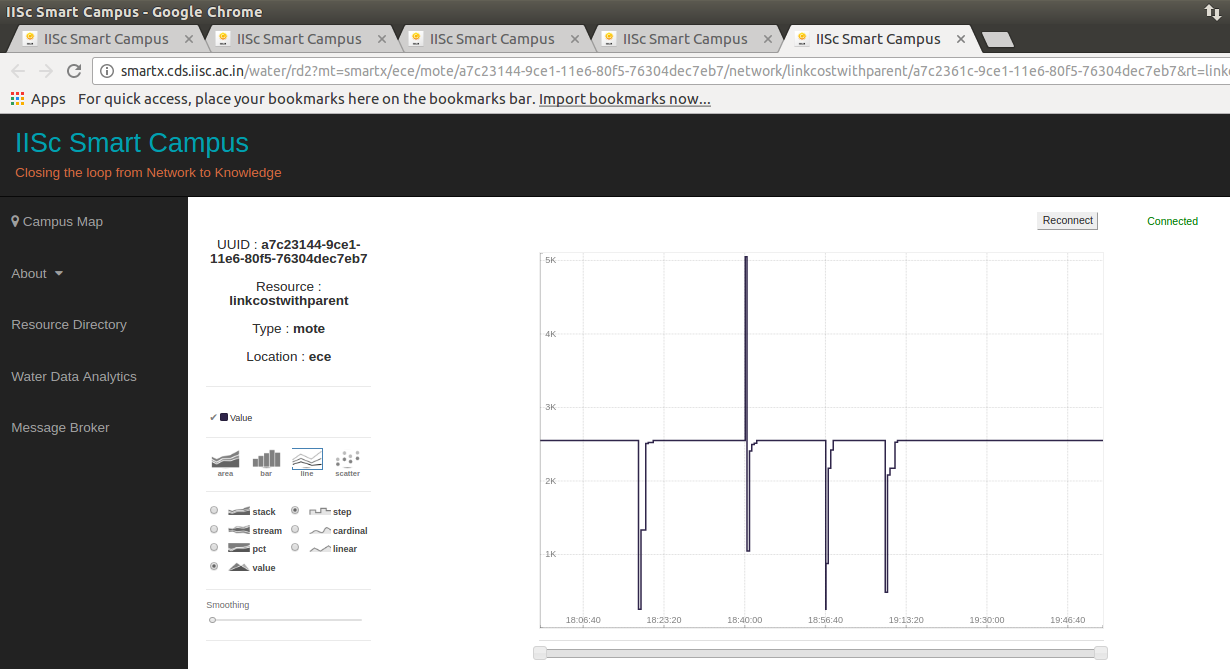}
  \caption{Real-time visualization of published observations}
    \label{fig:streamviz}
\end{figure}

Basic statistical analyses are performed over the cleaned data to offer a summarized view of the state of the IoT system. These are used for monitoring the domain or the IoT fabric, information dissemination across campus users, or for human decision making. The STATS streaming dataflow (Fig.~\ref{fig:pipelines}) performs operations like statistical aggregation, moving window averages, probability distributions, and basic plotting using libraries like XChart. Our STATS pipeline subscribes to the MQTT topic to which ETL publishes the cleaned observation streams. The statistical aggregates generated by STATS are likewise published to MQTT from which, e.g., the portal can plot realtime visualizations like Fig.~\ref{fig:streamviz}, while the plotted files are pushed to file store which can then be embedded in static webpages or reports.

Earlier, we have developed the \emph{RIoTBench benchmark} that has composable IoT logic blocks and generic IoT dataflows that are used for evaluating DSPS platforms~\cite{shukla:riot:2017}. We customize and configure these dataflow pipelines for the Smart Campus and the water management domain. As a validation of the scalability of the proposed solution, we have shown that Apache Storm can support event rates of over $1000/sec$ for many classes of tasks, as illustrated in Fig.~\ref{fig:stream:perf:storm}, when operating on an Azure Cloud VM. The tasks that were benchmarked span different IoT business logic categories such as parsing sensor payloads, filtering and quality checks, statistical and predictive analytics, and Cloud I/O operations. These are then assembled together and customized for the domain processing analytics, such as smart water management.

\modc{While we use Storm as our preferred DSPS in our software stack, it can be transparently replaced by any other DSPS that can compose these dataflow pipelines. The tasks we leverage from RIoTBench are designed as Java libraries, and hence many stream processing systems can incorporate them directly for modular composition. Since the interaction between these pipelines is through the MQTT pub-sub broker, it offers loose coupling between the dataflows and the platform components. In fact, multiple DSPS can co-exist if need be, say, to support higher-order queries using Spark or a lambda-architecture over streaming and static data using Flink. Even our Apollo MQTT broker can be replaced by Cloud-based pub-sub platforms like Azure IoT Hub that uses the MQTT protocol. Likewise, our choice of HBase can be replaced by other NoSQL platforms or Cloud Storage like Azure Tables as well. As we note next in \S~\ref{sec:analyze}, the HDFS or NoSQL store plays a similar role of loose coupling between Big Data batch processing platforms that need to operate on archived data.}

\section{Data Analytics and Decision Making}
\label{sec:analyze}
\begin{figure}[t]
	\centering
\subfloat[\modc{Round Trip Latency}]{
    \includegraphics[width=0.3\columnwidth]{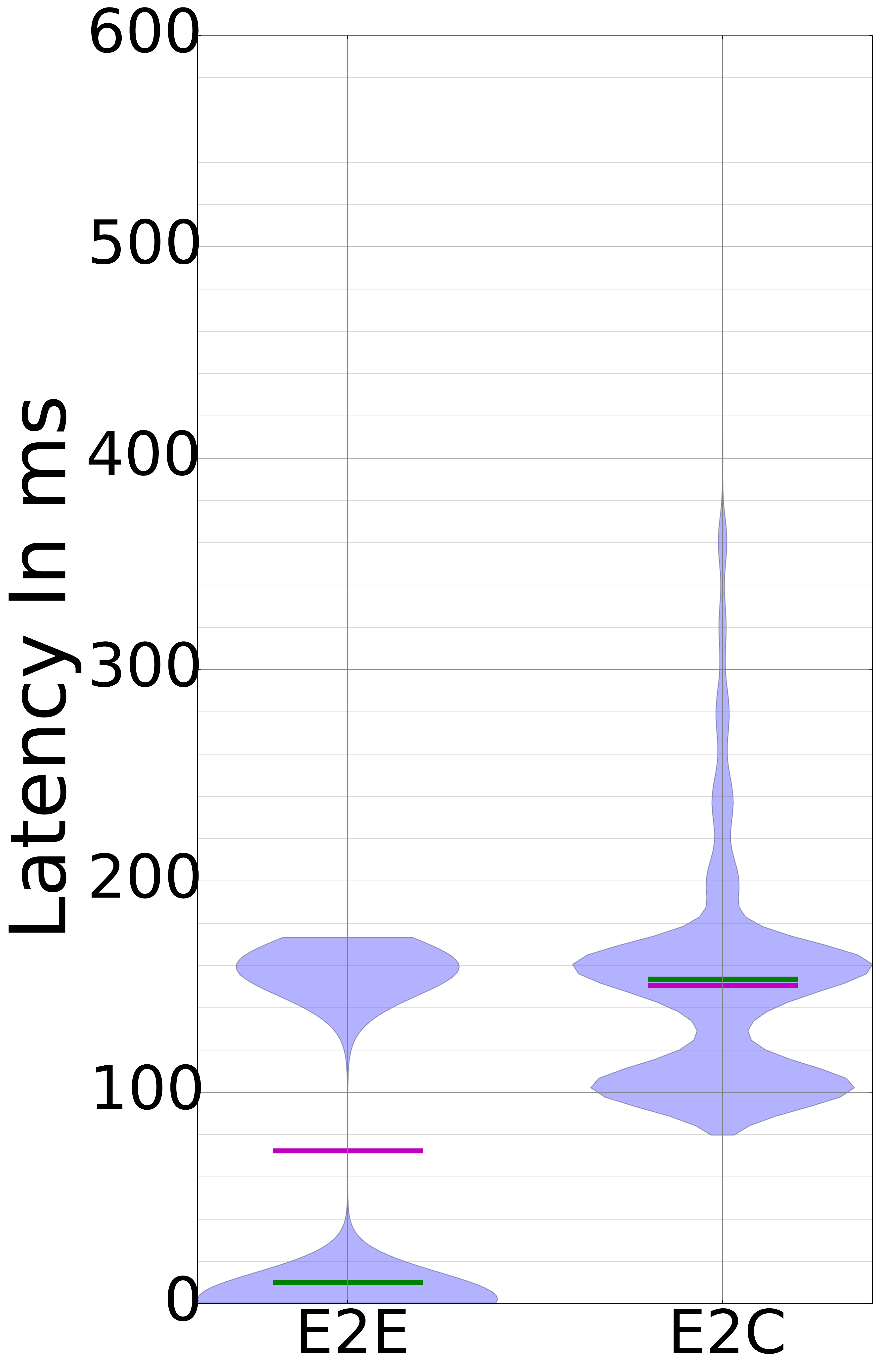}
  	\label{fig:perf:nw:lat}
  }~
  \subfloat[\modc{Network Bandwidth}]{
    \includegraphics[width=0.3\columnwidth]{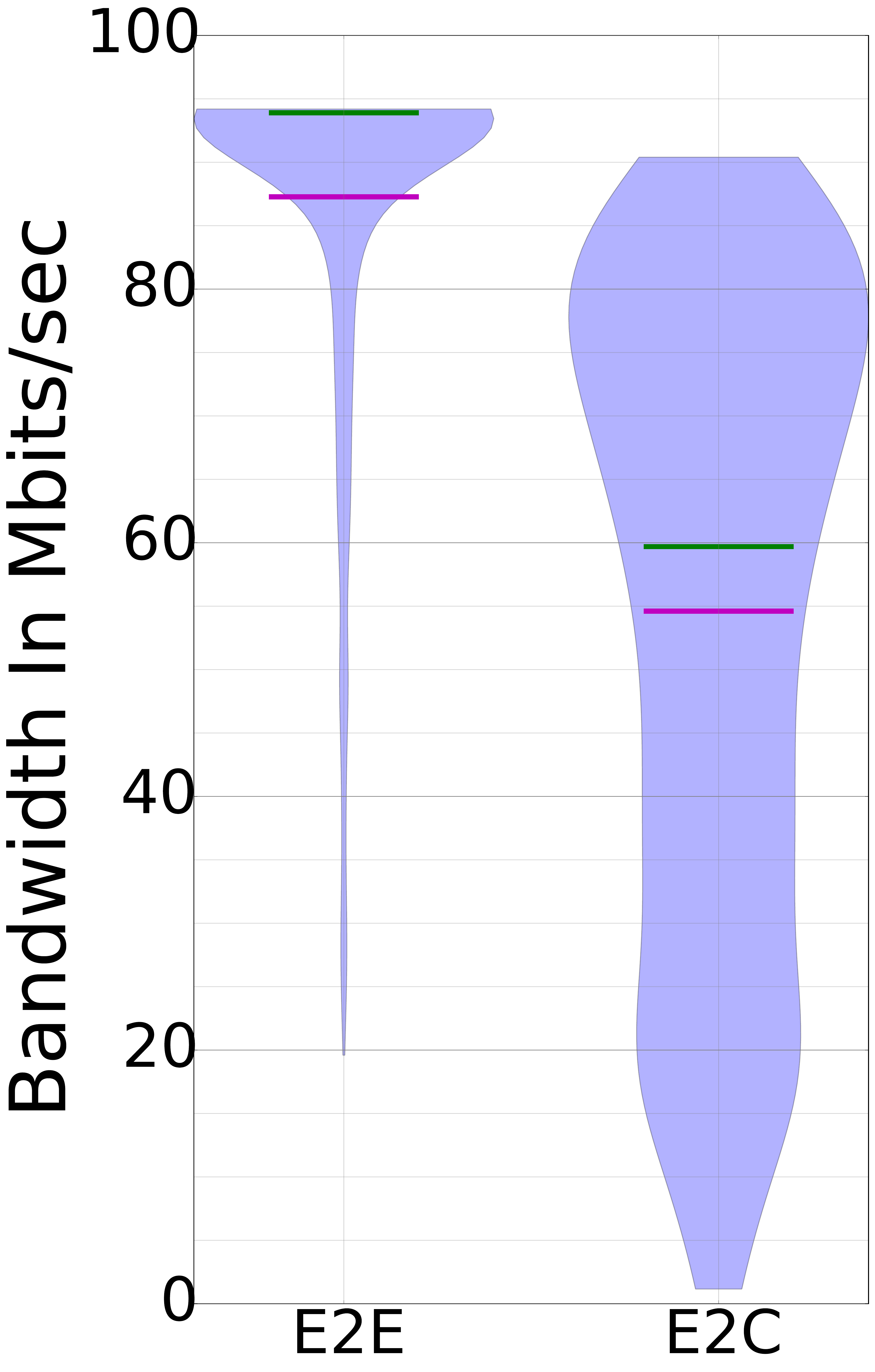}
  	\label{fig:perf:nw:bw}
  }\\
  \subfloat[Peak task input rate on an Azure VM~\cite{shukla:riot:2017}.]{
    \includegraphics[width=0.6\columnwidth]{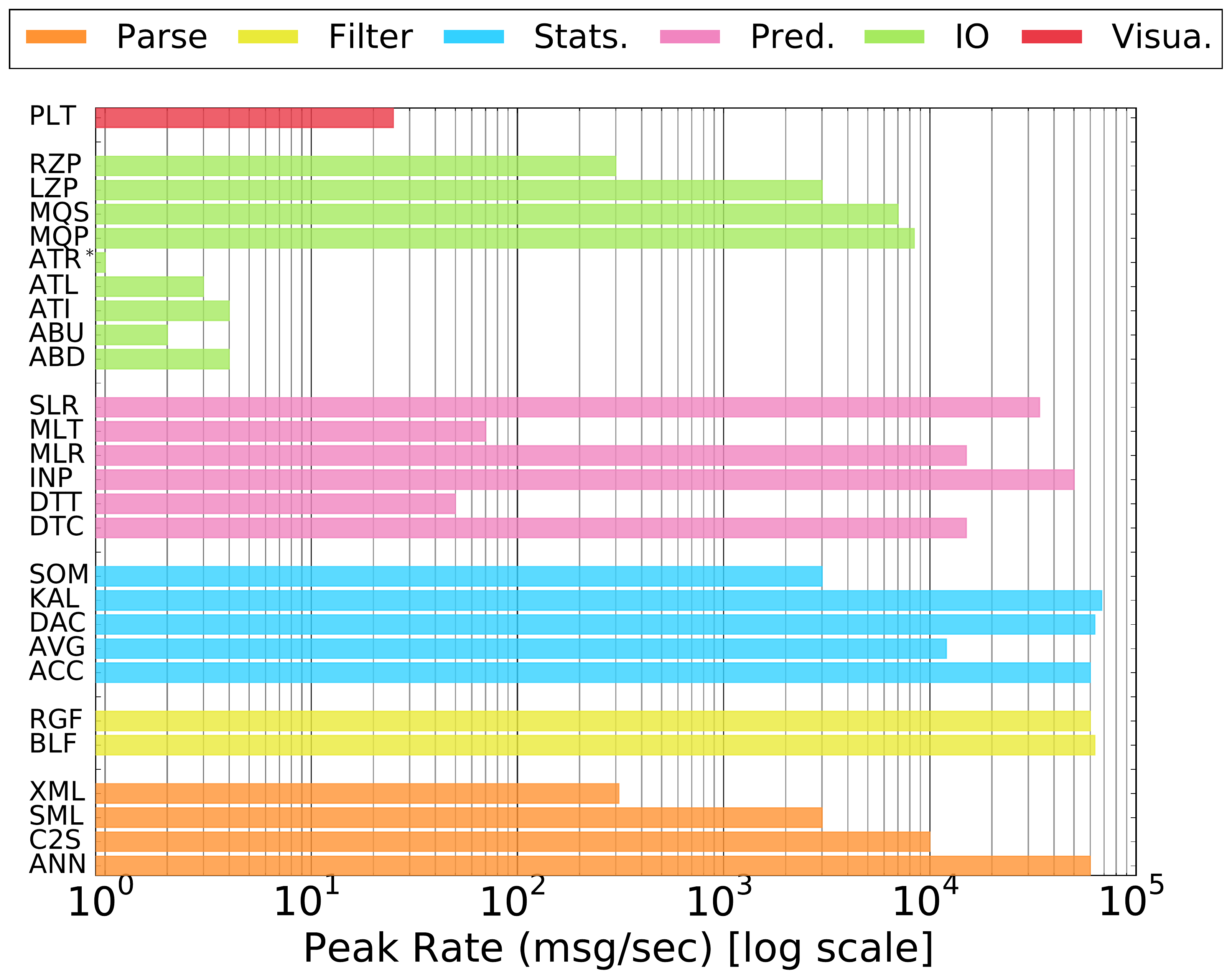}
  	\label{fig:stream:perf:storm}
  }
\caption{(a) Network latency and (b) Bandwidth distribution within Campus edge LAN (E2E) and from Campus to Cloud WAN (E2C). (c) Peak input stream rate supported for each Apache Storm DSPS task.}
\label{fig:stream:perf}
\end{figure}

\begin{figure}[t]
\centering
  \subfloat[Peak Query Throughput on Pi]{
    \includegraphics[width=0.45\columnwidth]{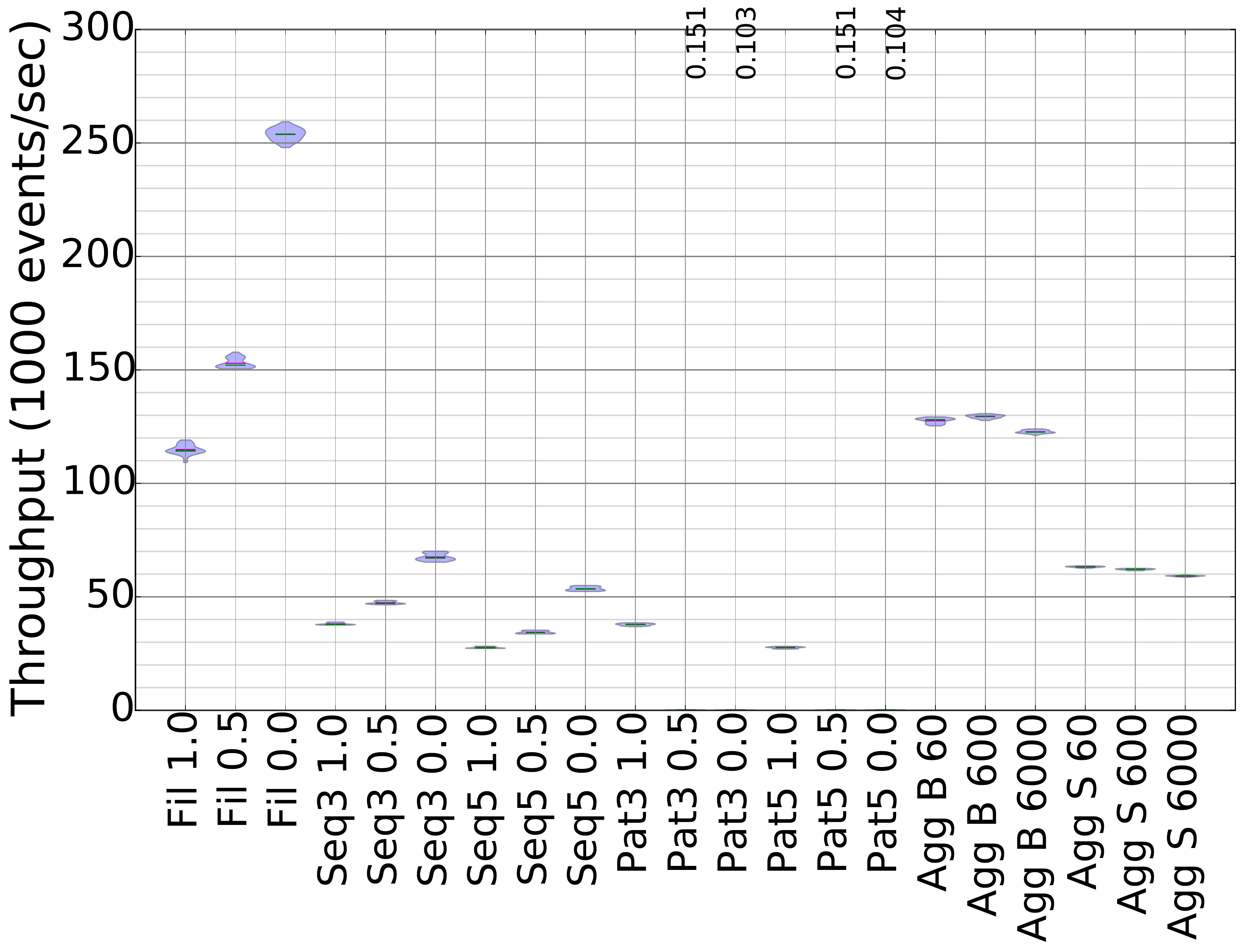}
  	\label{fig:cep:thruput:pi}
  }~~
  \subfloat[Peak Query Throughput on Azure]{
    \includegraphics[width=0.45\columnwidth]{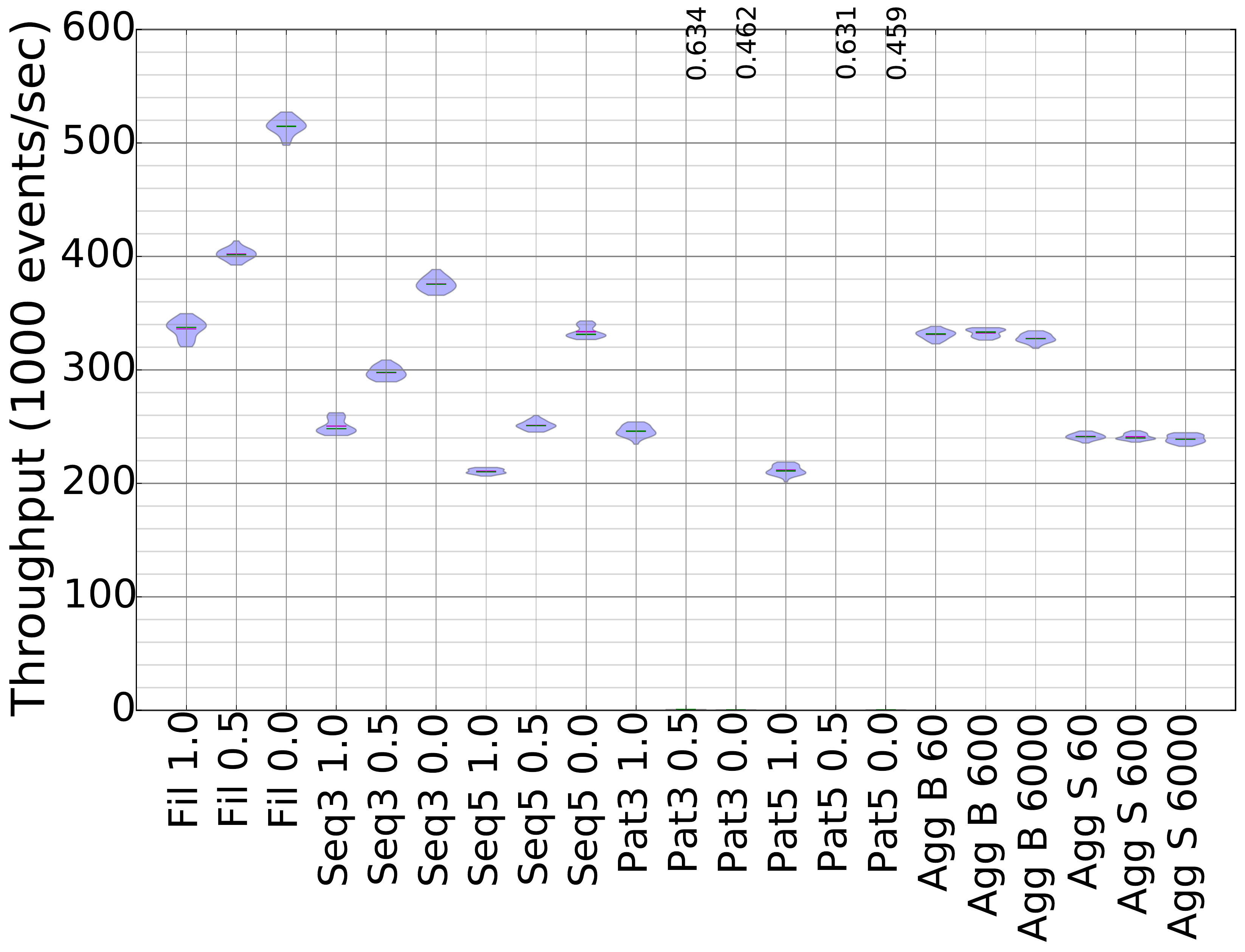}
  	\label{fig:cep:thruput:az}
  }\\   
   \subfloat[\modc{Query Latency on Pi}]{
    \includegraphics[width=0.45\columnwidth]{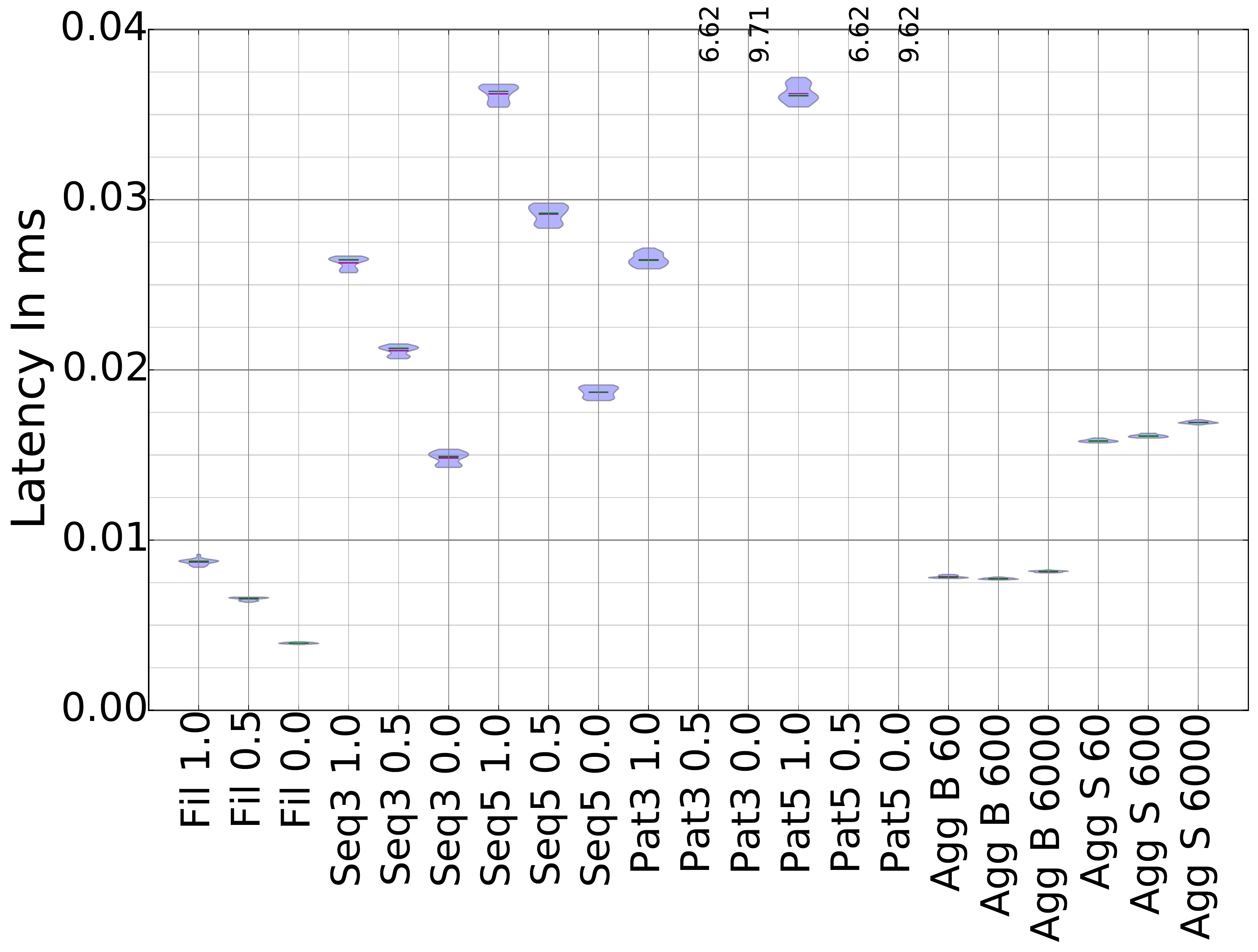}
  	\label{fig:cep:lat:pi}
  }~~\subfloat[\modc{Query Latency on Azure}]{
    \includegraphics[width=0.45\columnwidth]{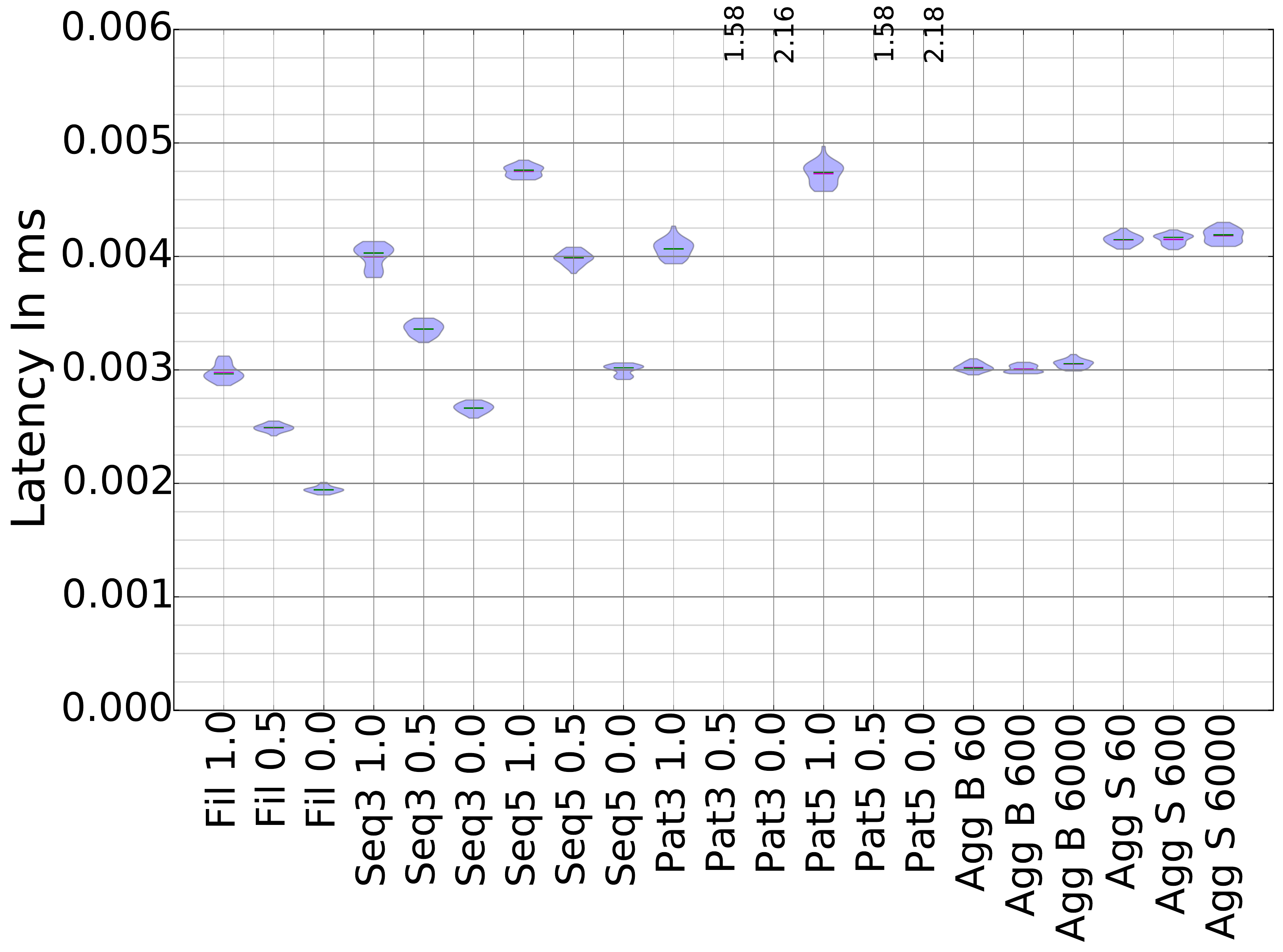}
  	\label{fig:cep:lat:az}
  }
\caption{\emph{Peak Throughput} and respective \emph{Query Latency} for various CEP queries on Pi and Azure VM~\cite{ghosh:tcps:2017}}
\label{fig:stream:perf:cep}
\end{figure}

There are several types of analytics that can help with manual and automated decision-making about the water domain, and the IoT fabric management as well. Similar to the stream processing pipelines for data acquisition above, streaming dataflows can perform analytics and decision making as well. Fig.~\ref{fig:pipelines} shows such \emph{online analytics and decision making pipeline (DECIDE)} that consumes cleaned observation streams from MQTT and can perform time-series analysis using auto-regressive models for, say, water demand prediction. Feature-based analytics, such as decision tree, can be embedded to correlate environmental observations with specific outcomes, such as days of the week with the water footprint in buildings. The figure also shows how such predictive models can be trained using streaming or batch dataflows from historic data (\emph{TRAIN}), and the updated models feed into the online predictions, periodically. We use logic blocks from the \emph{Weka} library~\cite{weka} within the Apache Storm dataflow for such online \emph{predictive analytics}, and several are made available as part of RIoTBench.

One of the most intuitive analytics for utility management is through the detection of event patterns. \emph{Complex Event Processing (CEP)} enables a form of \emph{reactive analytics} by allowing us to specify patterns over event streams, and identify situations of interest~\cite{cep}. It uses a query model similar to SQL that executes continuously over the event stream, and specifically allows window aggregations and sequence matching. The former applies an aggregation function over count or time windows, in a batch or sliding manner, while the latter allows a sequence of events matching specific predicates to be detected. E.g., these queries can detect when the moving average of water pressure goes above a certain threshold, or when the water level in a tank drops by $>5\%$ over successive events spread over $10~mins$. The former may indicate a blockage in the water distribution network, while the latter may identify rapid water leakage in building~\cite{govindarajan:comad:2014}. 

We use \emph{WSO2 Siddhi}~\cite{siddhi} as our CEP engine for such ``fast data'' event-analytics and validate its scalability both on gateway devices such as a Raspberry Pi 2 for edge-computing, as well as on an Azure VM for Cloud computing~\cite{ghosh:tcps:2017}. Fig.~\ref{fig:stream:perf:cep} shows prior results for $21$ representative queries that perform sequence and pattern matching, filtering and aggregation, etc. over water level streams on the Pi. As we can see, these event queries are light-weight and can support rates of over $25,000~events/sec$ even on a Pi, with the corresponding Azure benchmarks showing a $3\times$ improvement (Figs.~\ref{fig:cep:thruput:pi} and ~\ref{fig:cep:thruput:az}). \modc{We can also infer the per-event query latency from these peak throughputs (Figs.~\ref{fig:cep:lat:pi} and ~\ref{fig:cep:lat:az}), and most execute in $\le 0.04~ms$ on the Pi and in $\le 0.005~ms$ on the Cloud. There is limited variability in the execution latency or the throughput. While the execution on the Cloud is much faster, when coupled with the Edge-to-Cloud latency for transferring the event from a sensor on campus to the Cloud (Fig.~\ref{fig:perf:nw:lat}), execution on the Pi has a lower makespan.} These validate the use of event analytics for both edge and Cloud computing.

These analytics can provide trends, classifications, patterns, etc. that can then be used by humans to manually take decisions, or for rule-based systems to automatically enact controls. These actions can include 
automatically turning water-pumps on and off based on the water level, notifying users of contamination in a spatial water network region, reporting leaking pipes and taps to maintenance crew, etc. These strategies are currently being investigated as a meaningful corpus of water distribution and usage data within the campus is accumulated. Computational and network models that leverage these sensed data are being developed by our collaborators as well~\cite{msmk}.

In addition, these data streams and analytics help more immediately with understanding and managing the IoT fabric, particularly during the development and deployment phase of the infrastructure. They help identify, say, when the WSN is unable to form a tree or has high packet drops, when the sensors and motes are going to drain their battery, or when gateways go offline (e.g., due to \emph{wild monkeys} fiddling with the devices, as we have seen!). It also helps validate the performance of network algorithms like RPL, and build a repository of network signal strengths at different parts of campus, over time. 

\begin{figure}[t]
	\centering
	\includegraphics[width=0.75\columnwidth]{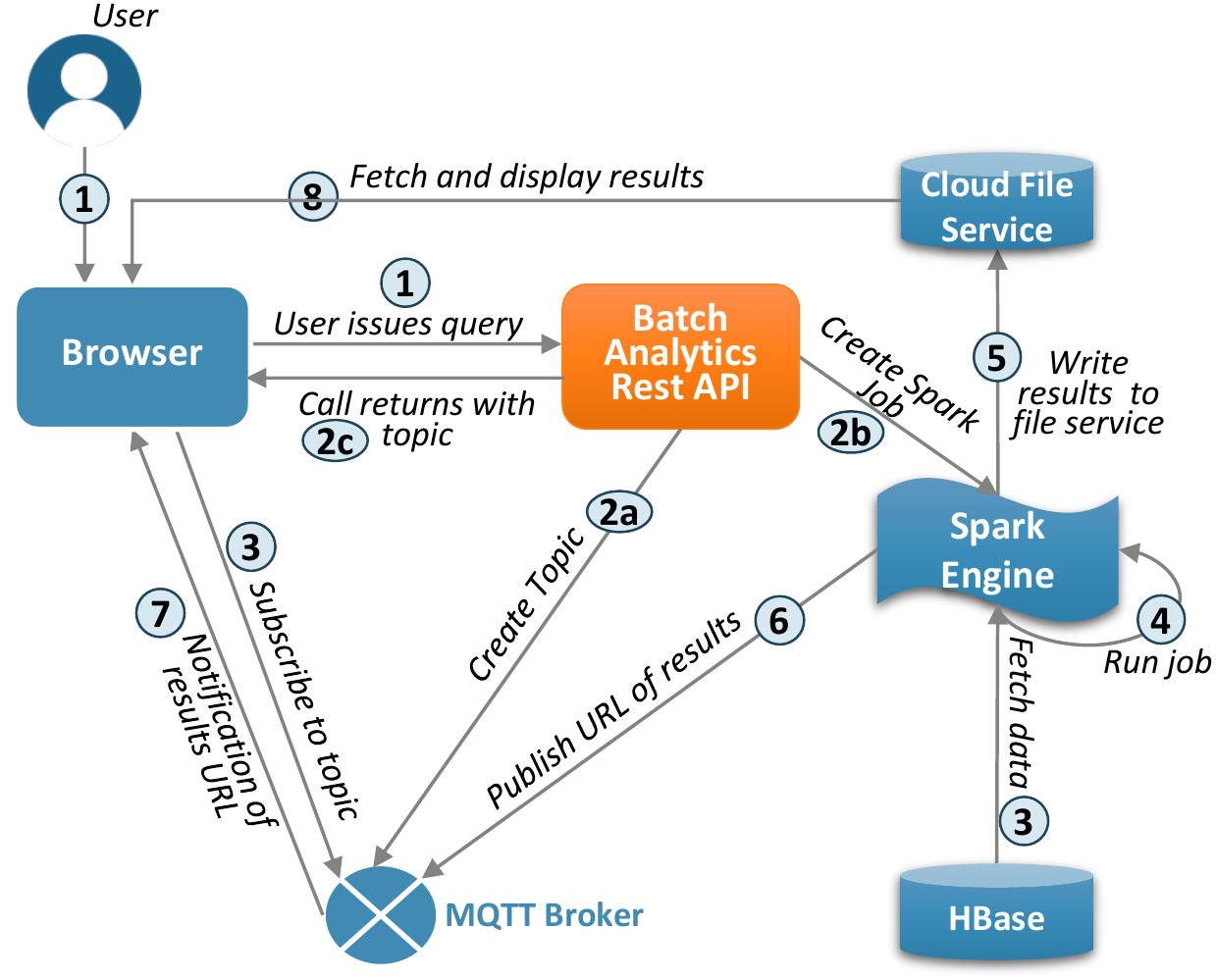}
	\caption{Workflow for Asynchronous Batch Analytics Service}
	\label{fig:batch}
\end{figure}

Often, these exploratory analyses are performed on historic data collected over days and months within our data platform.
We leverage the \emph{Apache Spark}~\cite{spark} distributed data processing engine for such batch analytics. Spark allows fast, in-memory iterative computations and has been shown to out-perform traditional Hadoop MapReduce platforms. It also offers intuitive programming models such as SparkQL for easy specification of analytics requirements. Spark uses HBase, where we archive the cleansed sensor data, as its distributed data source. It can also be used to train predictive models in batch using its Machine Learning libraries (MLLib). While we currently perform periodic model training using a Storm dataflow for convenience (\emph{TRAIN} in Fig.~\ref{fig:pipelines}), we propose to switch to Spark in the near future.

We expose a \emph{Batch Analytics REST Service} wrapper around Spark to ease the execution of simple analytics from the Smart Campus web portal. This allows temporal and sensor-based filtering, and aggregation and transformation operations over the observational datasets to be mapped as parameterized Spark jobs that run on Cloud VMs. The Spark jobs can run for several minutes to hours, depending on the complexity of the analysis and source data size, and generate KB to GB of data. Hence, a synchronous REST call from the portal will timeout. Instead, we define an asynchronous service pattern based on the existing architectural components, as shown in Fig.~\ref{fig:batch}. 

When the user submits an analytics query from their browser, the REST service first creates a unique MQTT topic for this session, and then invokes a Spark job by populating its parameters, including this topic. The REST service returns this topic to the browser, which subscribes to the topic with the broker. The Spark engine fetches the source data from HBase, runs the analysis, and writes the output to a Cloud file storage. It then publishes the URL of this result file to the unique topic in the broker. The browser gets notified of this URL and can use it to either stream and visualize the results, or allow the user to download it. This exhibits the flexibility of our service-oriented architecture to easily compose complex data management and analytics operations. \modc{In future, this REST API and asynchronous execution pattern can be easily extended to allow \emph{ad hoc} Spark SQL queries to be directly submitted for execution. This will allow developers to construct more powerful exploratory analytics, besides the user-oriented query template that is currently supported.}

\begin{figure}[t]
\centering
    \includegraphics[width=0.75\columnwidth]{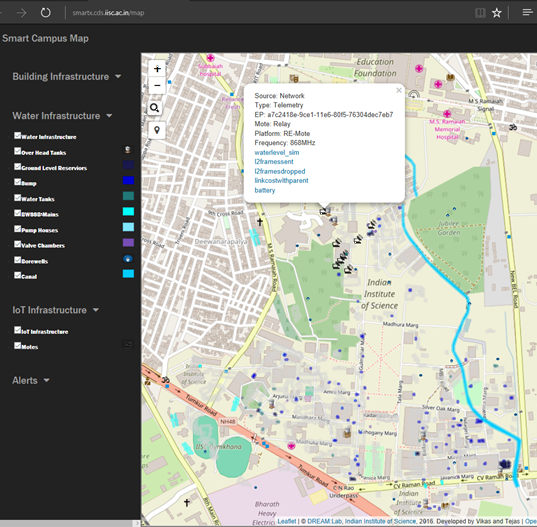}
  \caption{Geo-spatial visualization of Smart Campus water infrastructure, motes and sensors}
    \label{fig:viz}
\end{figure}
Lastly, we also support several types of \emph{visual analytics} \modc{that are
 exposed through the \emph{Smart Campus portal}. The portal itself was developed as part of this project, and includes a} dashboard for displaying real-time and temporal analytics (Fig.~\ref{fig:streamviz}) using JavaScript plugins like \texttt{D3.js} and \texttt{Rickshaw}, and also multi-layered geo-spatial visualizations of the IoT network on the IISc Campus using Open Street Maps (Fig.~\ref{fig:viz}). \modc{These leverage the self-describing SenML format used by the sensors for publishing observation streams, allowing plots to be automatically formatted for arbitrary sensors.}
These help with information dissemination to the end-users on campus, as well as simple visualization for resource managers. \modc{The portal also serves as a way for the campus managers to monitor the state of the IoT infrastructure using the RD, and potentially initiate actuation signals for enactment.}

\modc{As before for the choice DSPS, we can also replace Siddhi with other CEP engines like Apache Edgent, and Spark with platforms like Apache Pig or Hadoop. Our architectural design is agnostic to the specific platform, and the presence of pub-sub brokers and NoSQL data stores enable loose-coupling between diverse platforms that interface through them. Our selection of these specific platforms are indicative of what is adequate for the needs of the Smart Campus, and bounded by the scalability experiments that we have performed and reported. Other deployments may pick contemporary alternatives that are appropriate for their needs.}

\section{Related Work}
\label{sec:related}

There has been heightened interest recently in designing software fabrics and data platforms to manage IoT infrastructure, and data and applications within them, with even a special issue dedicated to such software systems~\cite{spe/ChenWZ17}. These are emerging from standards bodies \emph{(IETF CoRE, W3C Web of Things, ITU-T, ISO)}, industry and consortia \emph{(Azure IoT, AWS Greengrass, Threads Group, OneM2M, AllSeen Alliance, FIWARE, LoRa)}, and academia \emph{(IERC, IoT-A, OpenIoT)}, with implementations by the open source community \emph{(Californium, Kura, Sentilo, Kaa)}. While some of these, like MQTT, have gained traction, others are competing for mind-share and market share. However, we are at an early evolutionary stage and there is a lack of clarity on what would be the most suitable technical solutions, and what would gain popular acceptance (these being two different factors). In this context, having a practical implementation and validation of an integrated IoT architecture on the field using these functional designs and protocols, as we have presented in this article, will better inform these conceptual exercises and reference designs. While we make specific service-oriented design, protocol and implementation choices for the Smart Campus project, driven by Smart Utility needs in India, there are other numerous relevant efforts and alternatives, and we discuss a representative sample here.

\subsection{\modc{Community Specifications}}
The concept of a \emph{Web of Things (WoT)} was proposed several years back by W3C but did not translate to proactive standardization efforts like IETF's~\cite{guinard2011internet}. Recently, the W3C WoT working group has begun developing a formal WoT architecture for IoT~\cite{wot}. It leverages simplified forms of Web standards like HTTP, REST and JSON to support use-cases on Smart Homes, Smart Factory and Connected Cars. In this evolving draft, 
device, gateway (edge) and cloud are seen as first-class building blocks, similar to our own differentiation, and supported environments include browser, smart phones, edge hubs and cloud VMs. They also propose a \emph{servient} software stack to design and deploy applications built using a scripting framework, and protocol bindings to more popular IoT standards such as MQTT and CoAP. These bindings ensure that our own design that leverages existing standards is likely to be able to be able to interface with a WoT stack in the future.
The COMPOSE API for IoT~\cite{compose} takes a similar WoT view and defines REST operations and JSON payloads on Service Objects that wrap physical things. Applications can be composed across multiple physical devices using these APIs. 
While these are still early days for WoT, it is likely to find wider industry support given the past history of W3C standards.

\emph{OneM2M} is a broad-based effort to develop open specifications for a horizontal IoT middleware that will enable inter-operability for M2M communications. It proposes comprehensive service specifications for device identification, RESTful resource and container management, and synchronous and asynchronous information flows, with mappings to open protocols like CoAP, MQTT and HTTP~\cite{onem2m}. This is targeted at large-sale IoT deployments with complex devices and use-cases, and multiple vendors. This effort is driven by major telecom providers and government agencies such as US NIST and India's Department of Telecommunication, and is expected to gain traction once the standards are formalized. Our goal in this article is much more modest, and we validate a slice of these complex interactions within the campus-scale IoT deployment, using similar functional layers and open protocols.

\subsection{\modc{Open Source Efforts}}
\modc{\emph{FIWARE}~\cite{fiware}~\footnote{https://www.fiware.org} is an open IoT software standard and a platform that is gaining recent attention, and whose features overlap with our middleware requirements. Like us, it uses MQTT and CoAP protocols for accessing observations that are coordinated through a context broker, supports CEP processing using IBM PROTON for alerts, and uses HDFS and Hive for archival storage and querying. It pays particular attention to capturing the device context, data models, dashboarding and security, making it a holistic solution. However, our proposed architecture pushes this abstraction down to the network layer, with patterns for capturing data across public and private networks, and across embedded and gateway devices, as discussed in \S~\ref{sec:acquire}. We also place emphasis on the post-processing of captured event streams by DSPS to make them ready for analytics. These are practical needs from the field. That said, FIWARE can be used as a base implementation that is complemented with these mechanisms we propose.}

\emph{WSO2} has proposed a reference architecture for an IoT platform, much like IoT-A, as a starting point for software architects~\cite{wso2arch}. However, they focus primarily on the data and analytics platform rather than the networking and fabric management, which are essential on the field. They too leverage CoAP, MQTT and HTTP for communications, but unlike us abstract away device and communication concerns. They have their custom device management interface, targeted more at smart phones, and identity management for users using LDAP. They offer an open implementation of their WSO2 IoT software stack that supports MQTT message brokering, event analytics using Siddhi engine (which we also use), and enterprise dashboarding. Commercial software support is also offered.

\emph{Sentilo}~\cite{sentilo}~\footnote{https://www.sentilo.org/} is an open source platform for managing sensors and actuators in a smart city environment, supported by the Barcelona City Council. In their stack, devices need to be added to a catalog using a dashboard and a pre-defined data model to get an authentication token. Registered devices can then use their token to publish data and alerts to a Redis in-memory data store, that also has a pub-sub interface. Applications can register for these alerts and data changes and perform actions, but data pre-processing and analytics platforms nor their application logic are explicitly proposed. They also make no distinction between registered and online resources, unlike our LDAP and RD, and this makes it difficult to know the state of the devices without querying. They offer protocol adapters for SCADA and Smart Meters, but their design is not inherently suited for constrained devices. While it has similar architectural goals and functional elements as our design, it is not as grounded in standards compliance and inter-operability other than using RESTful APIs. However, they have deployed the stack at multiple city locations, giving it practical validation.

\subsection{\modc{Research Activities}} 
There are multiple efforts in the European Union (EU) on defining reference models for IoT, including IERC and AIOTI. \emph{Internet of Things-Architecture (IoT-A)} is one such EU FP7 project that proposes an application independent model that can then be mapped to a concrete architecture and platform-specific implementation~\cite{iota}. They offer a comprehensive survey of design requirements, and their reference architecture spans the device, communication, IoT service, virtual entity and business process layers, with service management and security serving as orthogonal layers. They also have a structured information model. This has a close correlation with our functional model, with device shadows (virtual entities) and security being gaps we need to address in the future. Further, rather than stop at a functional design, we also make specific platform and protocol choices for these functional entities, and deploy it in practice within the IISc campus. 

\modc{One of the key challenges of IoT networks and platforms is the plethora of co-existing and overlapping standards, and the need to interface across them. \emph{Aloi, et al.}~\cite{aloi2017enabling} highlight the need to operate over diverse communications technologies and network protocols as requirements for opportunistic IoT scenarios. Specifically, they examine the use of smart phones as mobile gateways to act as a bridge between communication protocols like ZigBee, Bluetooth, WiFi and 3G/4G. This abstracts the data access by the applications and user interface from the underlying technologies. Such a model is well suited for generalizing our crowd-sourced data collection using mobile apps, and offers a parallel with the sensor data management in our Pi gateways.}

\modc{Yet another dimension of large scale IoT deployments is the ability to plan the deployment ahead of time, and with limited field explorations. Here, modeling and simulation environments are useful design tools~\cite{chernyshev2017internet}. While our SmartConnect tool~\cite{smartconnect} helps with WSN design planning, more comprehensive tools exist to allow one to span sensing, networking, device management and data management design within the IoT ecosystem~\cite{fortino2017modeling}. Large scale deployments will benefit from mapping the proposed solutions to such simulation environments to evaluate specific technologies.}

A recent special journal issue focused on software systems to manage smart city applications that deal with large datasets~\cite{spe/ChenWZ17}. However, these articles fail to take a holistic view of the entire software stack and limited themselves to specific Big Data platforms such as Spark, or analytics techniques like Support Vector Machines (SVM). We instead investigate the fundamental software architecture design to support a wide variety of domain applications and analytics techniques.

\subsection{\modc{Smart City Deployments}} 
In this regard, other EU projects like \emph{OpenIoT} translate the IERC reference architecture into practical implementations~\cite{openiot}. However, they do not pay adequate attention to protocol choices for constrained devices and compatibility with emerging standards like CoRE, and offer just a proof-of-concept validation. The \emph{Ahab} framework goes further by examining the analytics stack that is necessitated by the use of both streaming and static smart city data through a lambda architecture~\cite{spe/VoglerSID17}. However, key aspects such as interaction models for device and sensor registration and the impact of network protocols are not considered.

\modc{The \emph{SmartSantander} testbed is one of the more progressive Smart City deployments, and it offers insights on traffic and human mobility from Spain~\cite{lanza2016managing,sanchez2014smartsantander}. They offer their design requirements, and a software architecture for managing the testbed. This includes gateway and server runtimes, registry services and resource management. Authentication, Authorization and Accounting (AAA) services, and sensor, actuator and application deployment through a service interface is provided as well. They offer examples of the potential data sources and analytics, such as environment monitoring, landscape irrigation, traffic and parking management. Many of our requirements and architectural design exhibit similarities.}

\section{Conclusion}
\label{sec:conclusion}

In this article, we have set out the design goals for an IoT fabric and data management platform in the context of Smart Utilities, with the IISc Campus serving as a testbed for validation and smart water management being the motivating domain. Our \emph{functional architecture} is similar to other IoT reference models, with layers for communication, data acquisition, analytics and decision making, and resource and device management. We also make \emph{specific protocol and software platform choices} that advance a data-driven, service-oriented design that integrates Big Data platforms and edge and Cloud computing. 
We also identify \emph{interaction patterns} for the integrated usage of these disparate standards, protocols and services that are evolving independently. At the same time, our design is \emph{generic to support other domains} such as smart power grids or intelligent transportation, and such a translation is underway as part of a ``lightpole computing'' effort within the Bangalore city~\cite{amrutur:iotdi:2017}. The experiences from the project will help in understanding the distinctive needs of Smart City utilities in developing countries like India.

Our performance results for the network design, as well as the Cloud-based stream pre-processing using Storm and edge-based event-analytics using Siddhi \emph{validate the scalability} of the software stack at the IISc campus. In particular, the platform is shown to scale to thousands of events per second for real IoT application logic on single VMs and Pi devices. These are inherently designed to weakly-scale, thus allowing these rates supported to further increase for city-wide deployments by adding more VMs and edge devices. The software stack also is available online as an open source contribution, allowing the open architecture design and implementation to be replicated at other campuses and communities as well. 

\modc{Having a service API and standards-based IoT middleware enables the rapid development of novel and practical applications, both for our intended goal of smart water management and beyond. Some such applications include mobile apps for crowd-sourced water quality reporting and user notification, with linkages to trouble-ticket management by the campus maintenance crew. These data sources are also helping with water balance study and leak detection applications within campus, such as ones done by our collaborators~\cite{amrutur:2016:cpsweek,msmk}. The key distinction is the ability to perform such studies on-demand and incorporate outcomes in real-time, rather than require custom time-consuming field experiments, as was the norm. This accelerates the translation of science into operational benefits. Further, the same IoT stack was used for crowd-sourced collection of WiFi signal strengths for use by the campus Information Technology team and for an IoT Summer School hackathon, as part of campus outreach programs~\cite{msr:school}.}

\modc{The initial field trials using hundreds of sensors and devices are underway across campus. However, to ensure that the scope of the project was kept manageable, several additional aspects were deferred for future exploration.} Key among them are security and policy frameworks which are essential in a public utility infrastructure~\cite{Bertino:2016:toit}. Several authentication and authorization standards already exist for the web, with billions of mobile devices and web application utilizing them. Utilities however have a higher threat perception and end-to-end security mechanisms will need to be enforced. Similarly, auditing and provenance will be essential to identify the operational decision making chain, especially with automation of mission-critical systems~\cite{Simmhan:ijwsr:2008}. Trust mechanisms have to be established for using crowd-sourced data for operations, and privacy within pervasive sensing is a concern. From a platform perspective, we are also investigating the use of edge and fog devices to complement a Cloud-centric data platform~\cite{prateeksha:icfec:2017,ghosh:tcps:2017,pushkar:icsoc:2017}. Energy aware computing and mobility of devices also needs attention. These will find place in our future work.

\bibliographystyle{plain}

\section{Acknowledgments}
This work was supported by grants from the \emph{Ministry of Electronics and Information Technology (MeitY), Government of India}; the \emph{Robert Bosch Center for Cyber Physical Systems (RBCCPS) at IISc}; Microsoft's \emph{Azure for Research} program; and \emph{VMWare}.

The authors acknowledge the contributions of other project investigators, M.S. Mohankumar, B. Amrutur and R. Sundaresan, to the design discussions and deployment activities. 

We also recognize the design and development efforts of other staff and students during the course of this project, including Abhilash K., Akshay P.M., Anand S.V.R., Anshu S., Arun V., Ashish J., Ashutosh S., Jay W., Jayanth K., Lovelesh P., Nithin J., Nithya G., Parama P., Prasant M., Ranjitha P., Rajrup G., Sieglinde P., Shashank S., Siva Prakash K.R., Tejus D.H., Vasanth R., Vikas H., Vyshak G., and  among others.

\bibliography{article}

\end{document}